\documentclass[twocolumn, trackchanges]{aastex631}
\usepackage{xcolor}
\usepackage{CJK}
\usepackage{apjfonts}

\newcommand\msun{$\rm M_{\odot}$}
\usepackage{ulem}
\usepackage{amsmath}
\usepackage{cancel}
\usepackage{soul}
\usepackage{subfigure}
\newcommand{\N}[1]{\textcolor{teal}{ #1}}

\usepackage{float}  
\usepackage{longtable}
\usepackage[utf8]{inputenc}
\usepackage[T1]{fontenc}

\def\msun{\hbox{M$_{\odot}$}}
\def\zsun{\hbox{Z$_{\odot}$}}

\def\OI{\ion{O}{1}} 
\def\HeI{\ion{He}{1}} 
\def\CaII{\ion{Ca}{2}} 
\def\Ha{H$\alpha$}
\def\Hb{H$\beta$}

\def\Msun{\(M_\odot\)}
\def\MgI{\ion{Mg}{1}}

\def\FeII{\ion{Fe}{2}}

\def\ScII{\ion{Sc}{2}}
\def\BaII{\ion{Ba}{2}}
\def\SiII{\ion{Si}{2}}
\def\NII{\ion{N}{2}}
\def\SII{\ion{S}{2}}

\shorttitle{SN~2023rve}
\shortauthors{Rosowsky et al.}

\begin{document}
\begin{CJK*}{UTF8}{gbsn}

\title{SN 2023rve: A Type II Supernova with No Nebular Oxygen}

\correspondingauthor{Melissa Rosowsky}
\email{mrrosowsky@ucdavis.edu}

\author[0009-0001-2376-0972]{Melissa Rosowsky}
\affiliation{Department of Physics and Astronomy, University of California, 1 Shields Avenue, Davis, CA 95616-5270, USA}

\author[0000-0001-8818-0795]{Stefano Valenti}
\affiliation{Department of Physics and Astronomy, University of California, 1 Shields Avenue, Davis, CA 95616-5270, USA}

\author[0000-0002-7352-7845]{Aravind P. Ravi}
\affiliation{Department of Physics and Astronomy, University of California, 1 Shields Avenue, Davis, CA 95616-5270, USA}

\author[0000-0002-7937-6371]{Yize Dong \begin{CJK*}{UTF8}{gbsn}(董一泽)\end{CJK*}}
\affiliation{Center for Astrophysics, Harvard \& Smithsonian, 60 Garden Street, Cambridge, MA 02138-1516, USA}

\author[0000-0002-7015-3446]{Nicolas Meza Retamal}
\affiliation{Department of Physics and Astronomy, University of California, 1 Shields Avenue, Davis, CA 95616-5270, USA}

\author[0000-0002-4924-444X]{K.\ Azalee Bostroem}
\altaffiliation{LSST-DA Catalyst Fellow}
\affiliation{Steward Observatory, University of Arizona, 933 North Cherry Avenue, Rm. N204, Tucson, AZ 85721-0065, USA}

\author[0000-0003-4102-380X]{David J.\ Sand}
\affiliation{Steward Observatory, University of Arizona, 933 North Cherry Avenue, Rm. N204, Tucson, AZ 85721-0065, USA}

\author[0000-0002-0832-2974]{Griffin Hosseinzadeh}
\affiliation{Department of Astronomy \& Astrophysics, University of California, San Diego, 9500 Gilman Drive, MC 0424, La Jolla, CA 92093-0424, USA}

\author[0000-0002-0744-0047]{Jeniveve Pearson}
\affiliation{Steward Observatory, University of Arizona, 933 North Cherry Avenue, Rm. N204, Tucson, AZ 85721-0065, USA}

\author[0000-0001-8005-4030]{Anders Jerkstrand}
\affiliation{The Oskar Klein Centre, Department of Astronomy, Stockholm University, AlbaNova, SE-10691 Stockholm, Sweden}

\author[0009-0008-9693-4348]{Darshana Mehta}
\affiliation{Department of Physics and Astronomy, University of California, 1 Shields Avenue, Davis, CA 95616-5270, USA}

\author[0000-0003-0123-0062]{Jennifer E.\ Andrews}
\affiliation{Gemini Observatory, 670 North A`ohoku Place, Hilo, HI 96720-2700, USA}

\author[0000-0002-1895-6639]{Moira Andrews}
\affiliation{Las Cumbres Observatory, 6740 Cortona Drive, Suite 102, Goleta, CA 93117-5575, USA}
\affiliation{Department of Physics, University of California, Santa Barbara, CA 93106-9530, USA}

\author[0000-0003-4914-5625]{Joseph Farah}
\affiliation{Las Cumbres Observatory, 6740 Cortona Drive, Suite 102, Goleta, CA 93117-5575, USA}
\affiliation{Department of Physics, University of California, Santa Barbara, CA 93106-9530, USA}

\author[0000-0003-2744-4755]{Emily Hoang}
\affil{Department of Physics and Astronomy, University of California, 1 Shields Avenue, Davis, CA 95616-5270, USA}

\author[0000-0003-4253-656X]{D.\ Andrew Howell}
\affiliation{Las Cumbres Observatory, 6740 Cortona Drive, Suite 102, Goleta, CA 93117-5575, USA}
\affiliation{Department of Physics, University of California, Santa Barbara, CA 93106-9530, USA}

\author[0000-0003-0549-3281]{Daryl Janzen}
\affiliation{Department of Physics \& Engineering Physics, University of Saskatchewan, 116 Science Place, Saskatoon, SK S7N 5E2, Canada}

\author[0000-0001-8738-6011]{Saurabh W.\ Jha}
\affiliation{Department of Physics and Astronomy, Rutgers, the State University of New Jersey,\\136 Frelinghuysen Road, Piscataway, NJ 08854-8019, USA}

\author[0000-0001-9589-3793]{Michael Lundquist}
\affiliation{W.~M.~Keck Observatory, 65-1120 M\=amalahoa Highway, Kamuela, HI 96743-8431, USA}

\author[0000-0001-5807-7893]{Curtis McCully}
\affiliation{Las Cumbres Observatory, 6740 Cortona Drive, Suite 102, Goleta, CA 93117-5575, USA}

\author[0000-0001-9570-0584]{Megan Newsome}
\affiliation{Department of Astronomy, University of Texas at Austin, 2515 Speedway, Austin, TX 78712, USA}

\author[0000-0003-0209-9246]{Estefania Padilla Gonzalez}
\affiliation{Space Telescope Science Institute, 3700 San Martin Drive, Baltimore, MD 21218, USA}

\author[0000-0002-4022-1874]{Manisha Shrestha}
\affil{School of Physics and Astronomy, Monash University, Clayton Campus, VIC 3800, Australia}

\author[0000-0003-0794-5982]{Giacomo Terreran}
\affiliation{Adler Planetarium, 1300 S DuSable Lake Shore Dr, Chicago, IL 60605, USA}



\begin{abstract}

We report on multiband photometric and spectroscopic observations of SN\,2023rve, a nearby Type II supernova (SN II) discovered in galaxy NGC 1097 (D=15.4 $\pm$ 3.2 Mpc). Nearby SNe II provide constraints on late-stage evolution and progenitor mass loss, particularly the role of circumstellar material (CSM) in shaping SN II observables. SN\,2023rve peaks with an absolute V-band magnitude of -17.1 and declines at a rate of 0.90 $\pm$ 0.02 mag/50 days during the plateau. The bolometric light curve implies a $^{56}$Ni mass of 0.0064 \(M_\odot\). Using hydrodynamic light-curve modeling, we infer an intermediate-mass progenitor ($\sim14-18 M_\odot$), a low explosion energy of 0.27$\times10^{51}$ ergs, and a dense CSM component with radial extent of 2900 \(R_\odot\) and density parameter of $10^{18}$ g cm$^{-1}$. This supports growing evidence that enhanced pre-SN mass loss influences the diversity of SNe II. The nebular spectra of SN\,2023rve show narrow \HeI{} lines and an absence of [\OI{}] lines unprecedented among Type II SNe. Comparison with other SNe II shows that only two other known objects, both with higher velocities, lack oxygen signatures at similar epochs, <10\% of the sample. The lack of oxygen emission combined with low explosion energy, a long plateau, and a small synthesized nickel mass may be consistent with partial fallback of material onto the compact remnant. We also discuss alternative explanations for the suppressed oxygen emission, including dust formation, oxygen-calcium mixing, and ongoing CSM interaction.

\end{abstract}


\section{Introduction}\label{sec:intro}

Type II supernovae (SNe), are a type of core-collapse supernova that originate from massive stars (masses greater than 8 \(M_\odot\)) that retain hydrogen at the moment of explosion\footnote{We exclude 
the interaction-dominated Type IIn subclass from this discussion.}.
These events can be classified by the shape of their light curve and the spectral features present \citep{arcavietal2012}. In particular, Type IIP supernovae (SNe IIP) are characterized by a prolonged plateau phase lasting $\sim$80-120 days, during which the luminosity remains nearly constant as the ionized hydrogen recombines, before sharply declining into the radioactive decay-driven nebular phase. Type IIL SNe show a more linear decline over this timescale \citep{barbonetal1979,Filippenko1997,faran2014}.
Studies suggest that SNe II span a continuum of properties in both their light curves and spectral evolution, likely reflecting a range of progenitor characteristics and mass-loss histories \citep{anderson2014characterizing,Valenti2016}. This continuity is supported by hydrodynamical light-curve modeling \citep{Martinez_2022}, spectroscopic analyses \citep{gutierrez2017type, csornyeigutierrez2026}, and radiative-transfer models showing that differences in CSM interaction may partially explain the trends between IIP and IIL \citep{hillierdessart2019}.

\begin{figure}
\includegraphics[width=1\linewidth]{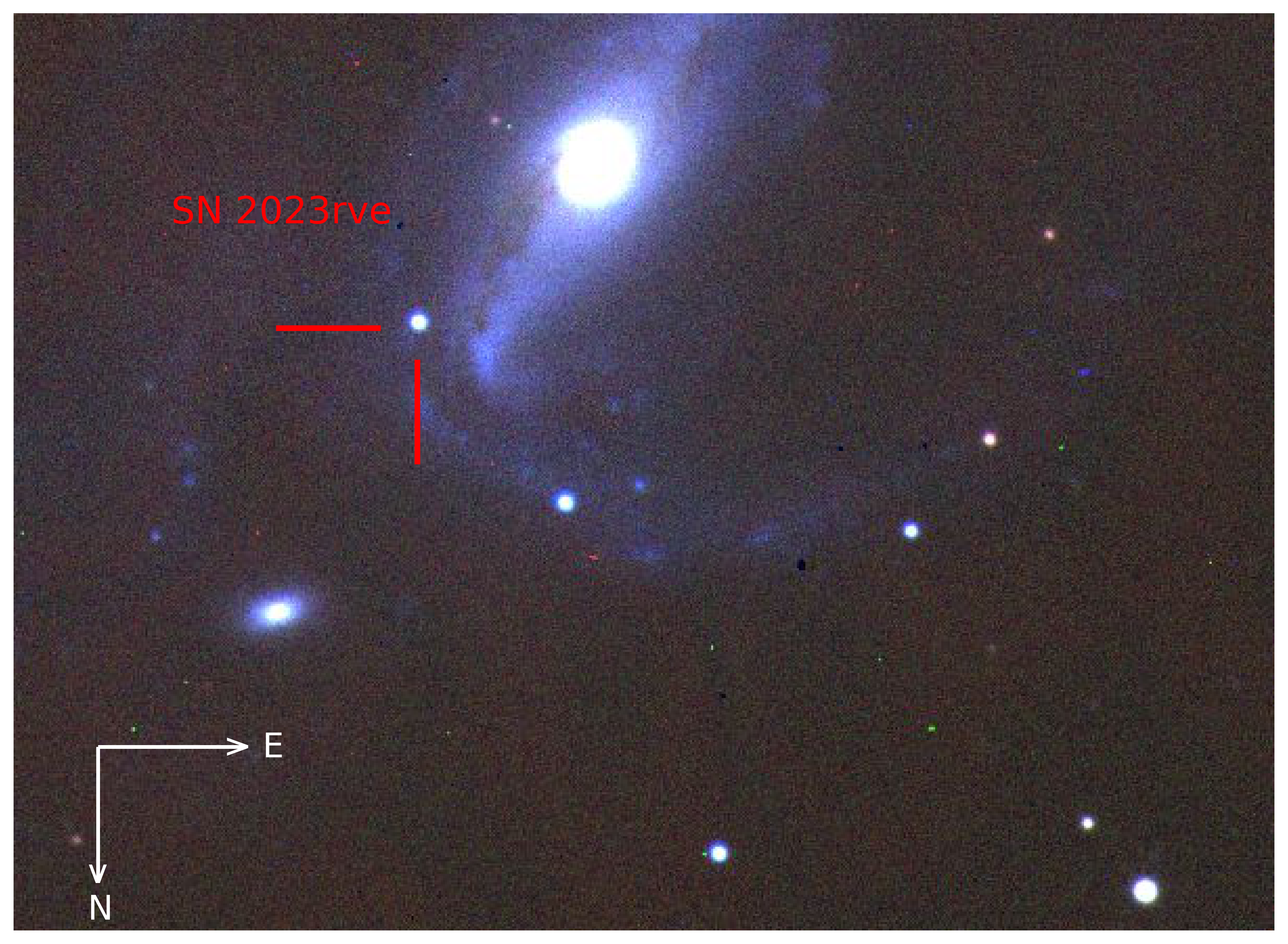}
\caption{RGB image of SN\,2023rve in galaxy NGC 1097, constructed using \textit{gri} images from the Las Cumbres Observatory (LCO). These images were taken September 14th, 2023, 10 days after the estimated explosion epoch of SN\,2023rve.
\label{fig:rgbimage}}
\end{figure}

Direct detections from pre-explosion images have confirmed that SNe II are typically associated with red supergiant (RSG) progenitors, placing most progenitor stars in the mass range $\sim$8-17 \(M_\odot\) \citep{smartt2009progenitors, smartt2015observational,daviesbeasor2020}. However, stellar evolution models predict that stars with initial masses up to $\sim$25 \(M_\odot\) should end their lives with enough hydrogen to produce SNe II, leading to a well-known discrepancy called the ``RSG problem'' \citep{smartt2009progenitors,smartt2015observational}. 
The significance of this discrepancy remains under debate. While \citet{2020MNRAS.493..468D} argued that uncertainties in progenitor mass estimates and the limited sample size weaken the evidence for a deficit of high-mass RSG progenitors, \citet{2020MNRAS.493.4945K} concluded that such a deficit remains significant even when adopting an alternative statistical framework.

This gap between the theoretical and observed mass range has led to a variety of proposed explanations. One possibility is that massive objects are enshrouded in dust, leading to lower estimates of luminosity and therefore mass \citep{beasordavies2016}. Another is that these objects could collapse undetected (with no visible explosion) to black holes, with fallback causing lower nickel mass and a less energetic explosion \citep{smartt2009progenitors}. It is also possible that observational bias against more massive progenitors contributes to this gap.

Some observed SNe II have been proposed as fallback explosions, where part of the ejecta falls back onto the compact remnant after explosion \citep{colgate1971, 2009ApJ...707..193F, zampieri2003}, producing low expansion velocities, faint radioactive tails, and potentially reduced nebular oxygen emission. Low-luminosity SNe IIP (LLSNe IIP) have been proposed as a class of such events \citep{zampieri2003}, though direct progenitor detections suggest they may instead arise from low-mass 8--10~\Msun{} progenitors \citep{smartt2009progenitors}, leaving the physical origin of these faint explosions an open question.

To resolve this discrepancy, several indirect techniques have been developed to estimate progenitor properties. Hydrodynamic modeling of SN light curves can yield estimates of ejecta mass, explosion energy, and progenitor radius \citep{morozova2017unifying,morozova2018measuring}, while late-time nebular spectral modeling can constrain the inner ejecta composition and infer the progenitor's zero-age main sequence (ZAMS) mass based on nucleosynthesis processes, particularly via oxygen emission \citep{Jerkstrand2012,jerkstrand2014nebular}.

Another open question for Type II SNe concerns the amount of mass lost by the progenitor star prior to explosion \citep{smith2014}. While RSGs are expected to lose mass through steady winds, recent observations suggest that many SNe II experience enhanced or episodic mass loss before core collapse \citep{Bruch_2021,smith2014,wufuller2021,jacobsongalan2022,Davies_2022}. 
Around 30\% of SNe~II show flash-ionization features indicative of shock breakout through a dense CSM \citep{forster2018, Bruch_2021}, as seen for example in the prototypical SN IIP SN\,2013fs \citep{Yaron2017}. The resulting CSM can alter the appearance of the transient, particularly in the days just after explosion \citep{chevalier1994}, and early-time spectroscopy provides some of the most direct evidence of this pre-SN mass loss: narrow emission or absorption features in the spectra arise from the interaction of the SN shock with nearby circumstellar gas (e.g.,\cite{galyam2014,singh2026}), and their strength and width can constrain both the mass and spatial distribution of the ejected material. Characterizing the CSM at early times, combined with nebular-phase spectroscopy (300--500 days after explosion) when the inner ejecta layers are revealed, provides a complete picture of the progenitor's mass-loss history and nucleosynthetic yields.


In this paper, we present spectroscopic and photometric data for SN\,2023rve, along with inferences about its progenitor. Progenitor information can be constrained using SN ejecta signatures through techniques like hydrodynamic modeling, nebular spectral modeling, and light curve comparisons. This paper is laid out with observations of SN\,2023rve in Section \ref{sec:observations}, reddening and host properties in Section \ref{sec:reddening_host}, and observational properties such as explosion epoch and distance in Section \ref{sec:obs_properties}. We describe the photometric evolution and present the pseudo-bolometric light curve of SN\,2023rve in Section \ref{sec:phot_evol}, and the spectroscopic evolution is shown in Section \ref{sec:spec}. We constrain the mass, energy, and CSM information in Section \ref{sec: prog_properties}, using hydrodynamic modeling. Finally, we list our conclusions in Section \ref{sec:conclusions}.

\begin{figure*}
    \centering
    \includegraphics[width = 1. \linewidth]{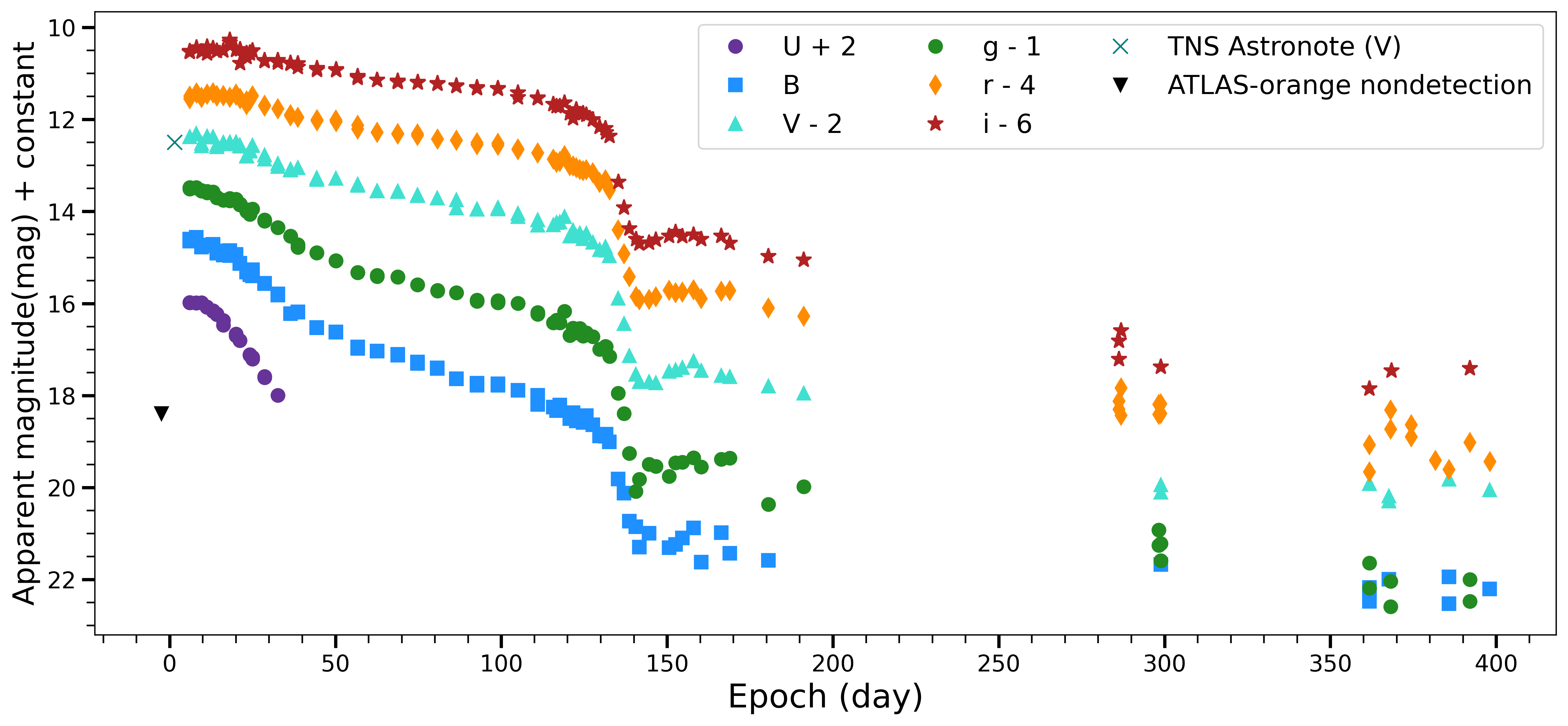}
    \caption{Multiband light curves for SN\,2023rve with respect to the assumed epoch of explosion (see \ref{tab:log_of_phot}). The pre-discovery imaging from Telescope Live \citep{astronote0906} provides a V-band magnitude, which is shown in addition to the last nondetection from ATLAS (in the orange filter).  
    \label{fig:lightcurve}}
\end{figure*}

\begin{deluxetable}{cc}
\tablecaption{Basic properties of SN\,2023rve used \label{tab:sn_properties}}
\tablewidth{0pt}
\tablehead{
\colhead{Property}&
\colhead{Value}
}
\startdata
Host galaxy & NGC~1097\\
RA (J2000) & 02\textsuperscript{hr}46\textsuperscript{m}18\fs13\\
DEC (2000) & $-30\degr 14\arcmin 22\farcs 16$\\
Distance & $15.4 \pm 3.2$ Mpc\\
Distance modulus& $30.94 \pm 0.45$ mag \\
Redshift $z$ & 0.004\\
$E(B-V)_{\rm MW}$ & $0.0231 \pm 0.0004$ mag$^\star$\\
$E(B-V)_{\rm host}$ & $0.13 \pm 0.03$ mag\\
Explosion epoch (JD) & $2460192.0 \pm 2.0$ (2023-09-04)\\
$V_{\rm max}$ & -17.11 $\pm 0.46$ mag (2023-09-14)\\
\enddata
\tablenotetext{\star}{\cite{Schlafly2011}.
}
\end{deluxetable}

\section{Observations}\label{sec:observations}
SN\,2023rve was discovered at RA (J2000) 2\textsuperscript{hr}46\textsuperscript{m}18\fs13 and Dec (J2000) -30\textsuperscript{hr}14\textsuperscript{m}22\fs16 in the galaxy NGC 1097 at an apparent magnitude of 14.4 mag (Figure \ref{fig:rgbimage}). It was discovered on September 8th, 2023 by Mohammad Odeh, using a 0.36m telescope through a supernova search program at the Al-Kahtim Astronomical Observatory in Abu Dhabi \citep{discovery}. Shortly after discovery, spectroscopic observations were conducted using the FOSC-ES32 spectrograph on a 0.41m Marcon telescope as part of the Italian Supernovae Search Program, which classified the object as a young Type II supernova \citep{classification}.

Prediscovery images from a Takahashi FSQ-106ED F3.6 telescope in Australia operated by Telescope Live dating back to September 6th, 2023 reveal the presence of SN\,2023rve in its early stages \citep{prediscovery}. ATLAS data places the last nondetection ($>$18.39 mag) on September 2nd, 2023. We constrain the explosion epoch by taking the middle of these dates, and therefore adopt 2023-09-04 (JD $2460192.0 \pm 2.0$) as the explosion epoch for the remainder of this paper.  

Soon after discovery, high-cadence observations using Las Cumbres Observatory were triggered as part of the Global Supernova Project (GSP) for 485 days, at which point the last nebular spectra was collected and the supernova fell below detection limits. Photometric data from the Las Cumbres network were reduced using the PyRAF-based photometric reduction pipeline \texttt{LCOGTSNPIPE} outlined in \cite{Valenti2016}, which utilizes a low-order polynomial and standard point-spread function (PSF) fitting method to remove background and estimate instrumental magnitudes. The pipeline \texttt{LCOGTSNPIPE} can also estimate apparent magnitude on galaxy-subtracted images.   
The apparent magnitude is then calibrated using APASS (B, V, \textit{g, r, i}) and Landolt (U) star catalogs. The resulting multi-band light curves are shown in Figure \ref{fig:lightcurve}, and a sample of the corresponding magnitudes are available in Table \ref{tab:log_of_phot}. 

\begin{figure*}
    \includegraphics[width=1.\linewidth]{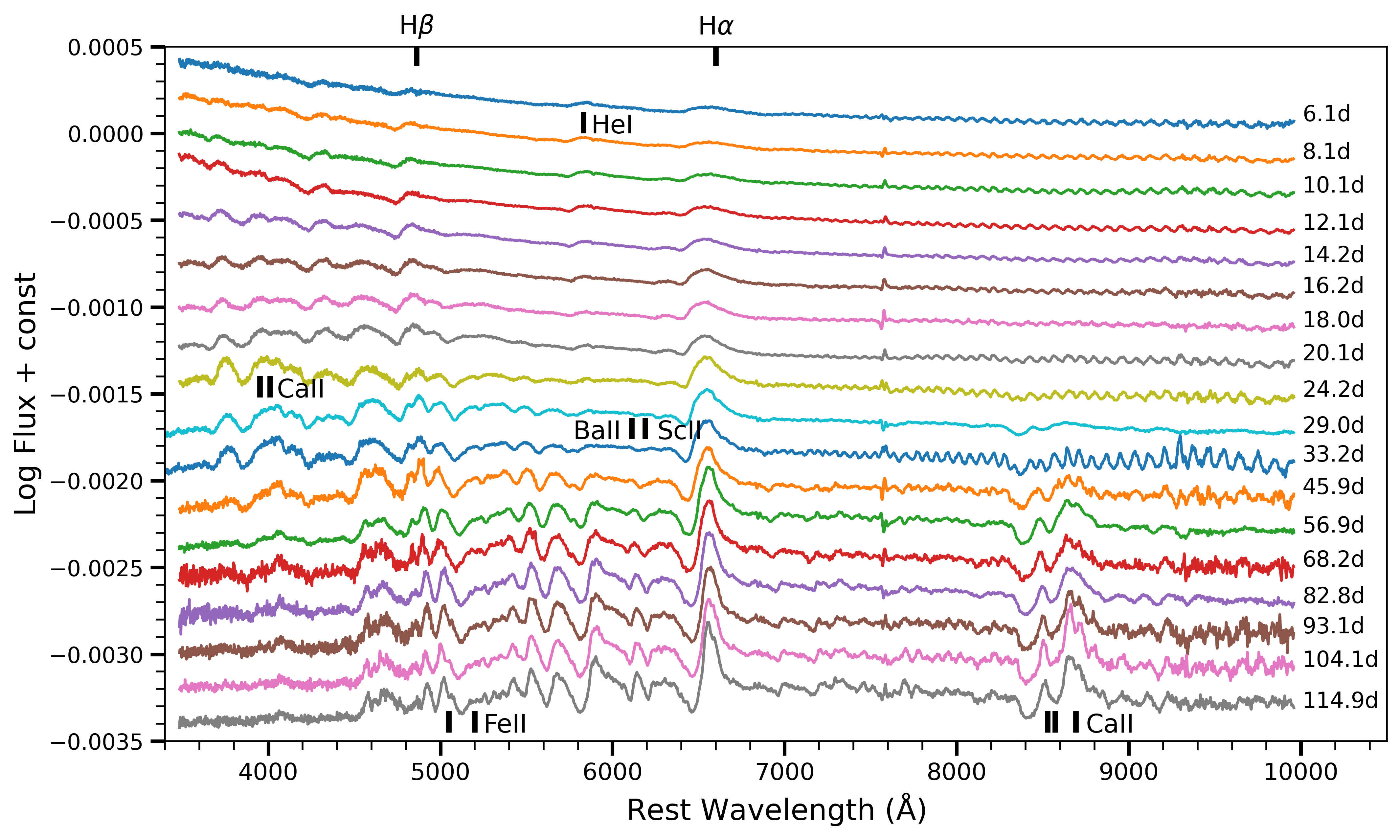}
    \caption{Spectroscopic evolution for SN\,2023rve from 6.1 days to 114.9 days after explosion.
\label{fig:evolution}}
\end{figure*}

Spectroscopic observations of SN\,2023rve began September 10th, 2023 and ended Januray 9th, 2025. The majority of the spectroscopic data was collected using the FLOYDS spectrographs \citep{Brown2013} mounted on the 2m Faulkes Telescope South (FTS), in Siding Spring, Australia, and Faulkes Telescope North (FTN), in Haleakala, using an approximately 1-day cadence for the month following discovery. The FLOYDS pipeline was used for the reduction of these spectra \citep{Valenti2014}. 
The pipeline splits the first- and second-order spectra into red and blue arms and rectifies them using a Legrendre Polynomial. Data are then trimmed, flat-fielded using images taken during the observing block and cleaned of cosmic rays. Red and blue arms are then flux and wavelength calibrated and then merged into a 1D spectrum using standard task in a python framework.
Two spectra were collected on September 19th and 21st with the Very Large Telescope in Chile (VLT), using the FORS2 spectrograph \citep{appenzeller1998successful}, which were reduced using an EsoReflex software \citep{freudling2013automated}. Three spectra were obtained from the LRIS instrument on the Keck Telescope on Mauna Kea at late-time during the nebular phase and were reduced using the LPIPE pipeline \citep{2019PASP..131h4503P}. A spectrum of SN 2023rve was also obtained with the RSS instrument on the Southern African Large Telescope (SALT) on UT 2025-01-09, but was dominated by host galaxy light.  A log of all spectroscopic observations is shown in Table \ref{tab:log_of_spec} and several spectra are plotted in Figure \ref{fig:evolution}.

\section{Reddening and Host Properties}\label{sec:reddening_host}
\subsection{Reddening}
The Milky Way line-of-sight reddening toward SN\,2023rve is $E(B-V)_\mathrm{MW} = 0.0231 \pm 0.0004$ mag, obtained from the Galactic dust maps of \cite{Schlafly2011}. Assuming a standard total to selective extinction ratio of $R_V = 3.1$, this corresponds to a V-band extinction of $A_V = 0.073$ mag.

The equivalent width of the NaID $\lambda\lambda$5890, 5896 absorption doublet can be used to estimate the host extinction due to its correlation with the amount of gas and dust present around the SN \citep{Poznanski2012}. Using a spectrum obtained on 2023-09-19 in which the two components of the doublet are unresolved, we measure a combined equivalent width of 0.881 \AA{} and apply the relation of \cite{Poznanski2012} for the blended feature, yielding $E(B-V)_\mathrm{host} = 0.13 \pm 0.03$.

\begin{figure}
\includegraphics[width=1.\linewidth]{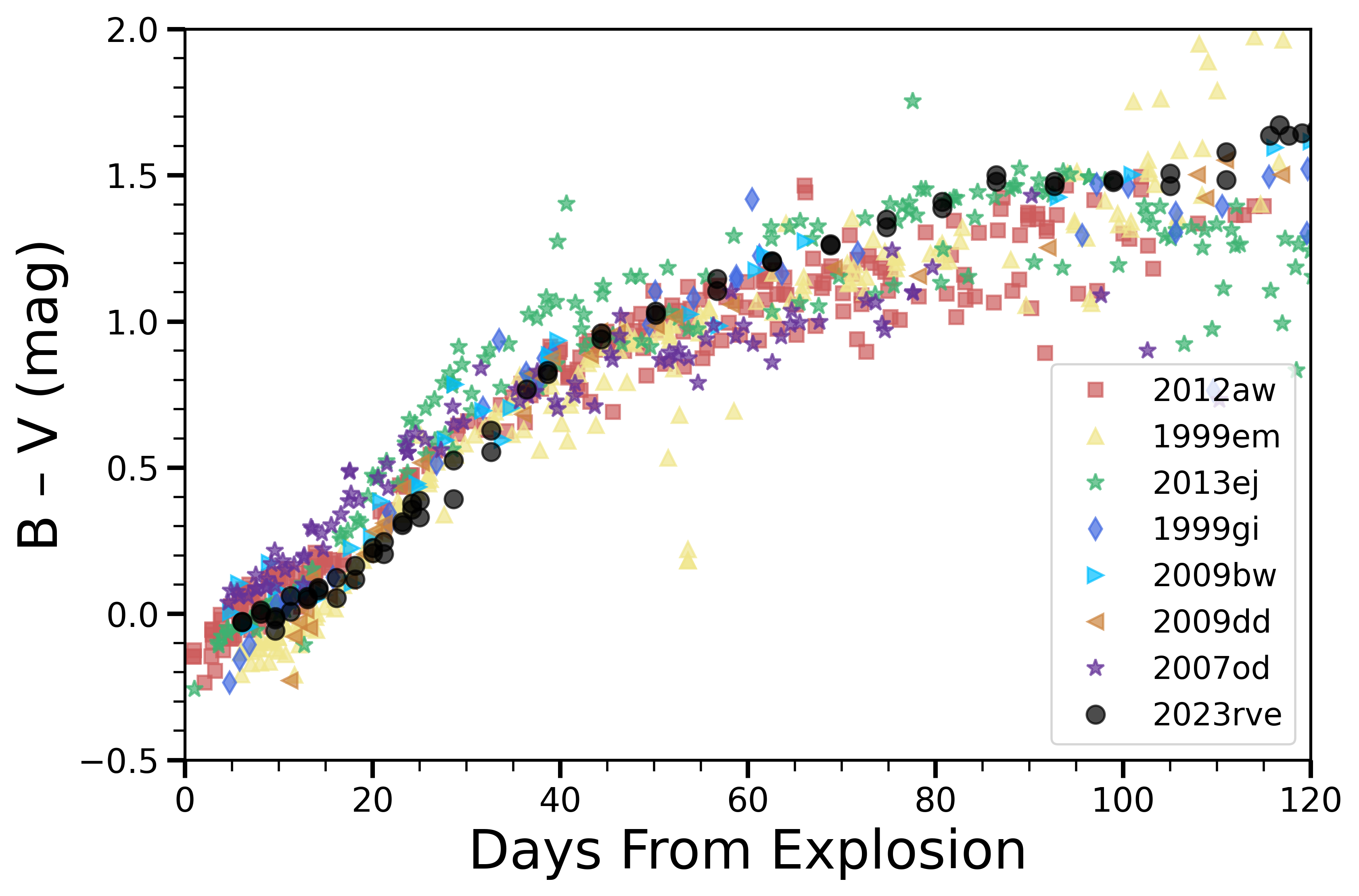}
\caption{Dereddened color evolution of SN\,2023rve and a sample of similar SNe II with published reddening values (see Section \ref{sec:reddening_host}). 
\label{fig:color}}
\end{figure}

To check the validity of this value, we compare the dereddened $B-V$ color evolution of SN\,2023rve with similar SNe II with known reddening estimates, utilizing the public Davis supernova database \footnote{https://dark.physics.ucdavis.edu/sndavis/transient}. The comparison sample used here and throughout the paper is composed of SN\,1999em \citep{Baron_2000}, SN\,1999gi \citep{1999gi}, SN\,2009bw \citep{2009bw}, SN\,2009dd \citep{2009dd}, SN\,2012aw \citep{dallora2014,bayless2013,Bose_2013}, SN\,2013ej \citep{Valenti_2013,2016MNRAS.461.2003Y}, and SN\,2007od \citep{2007odandrews}. Most of these objects have comparable slopes after maximum brightness in their light curves, which can indicate a similar color evolution according to \cite{de2018observed}. We include SN\,2007od in the sample, although it has a much earlier fall from plateau than SN\,2023rve, because it also has a comparable slope, and similarly does not seem to show forbidden oxygen lines at late time (see Section \ref{sec:spec}). The color comparison of this sample is shown in Figure \ref{fig:color}. The low reddening value suggested by the equivalent width measurement is supported by the color evolution compared to other objects with relatively low reddening. We maintain that the NaID-based result is a sufficient host reddening. Based on these results, we adopt a total reddening of $E(B-V)_\mathrm{tot} = 0.15 \pm 0.03$ for the remainder of this paper.

For extinction corrections in multiple photometric bands, we adopt the reddening law of \cite{cardelli1988determination} with $R_V = 3.1$ and apply it using the total reddening $E(B-V)_{\rm tot}$.

\subsection{Host Properties} 

 
NGC 1097 is a well-observed barred spiral galaxy in the constellation Fornax. The metallicity of the host galaxy environment can be constrained using features in the SN spectrum. In particular, \cite{dessart2014} showed that the strength of metal lines in Type II supernova spectra correlates with progenitor metallicity. Therefore, we measured pseudo-equivalent widths (pEWs) of the \FeII{} $\lambda5018$ feature to constrain the host metallicity in the location of the SN. We used a procedure in which each spectrum is first inspected visually, and the line region is then defined interactively by selecting wavelength boundaries that define the integration window. A local pseudo-continuum is estimated by fitting a first-order polynomial to narrow regions ($\pm10$ \AA) around the boundaries of the selected window. The spectrum is then normalized by this continuum fit, and the pEW is computed from this normalized spectrum.
Uncertainties on the pEW measurements were estimated empirically by repeating the measurement procedure for each spectrum. In each realization, the boundaries and continuum placement were re-selected, and the standard deviation of the resulting pEW values was adopted as the uncertainty. This approach captures the dominant systematic uncertainty associated with subjective continuum and boundary choice.

\begin{figure}
\includegraphics[width=1.\linewidth]{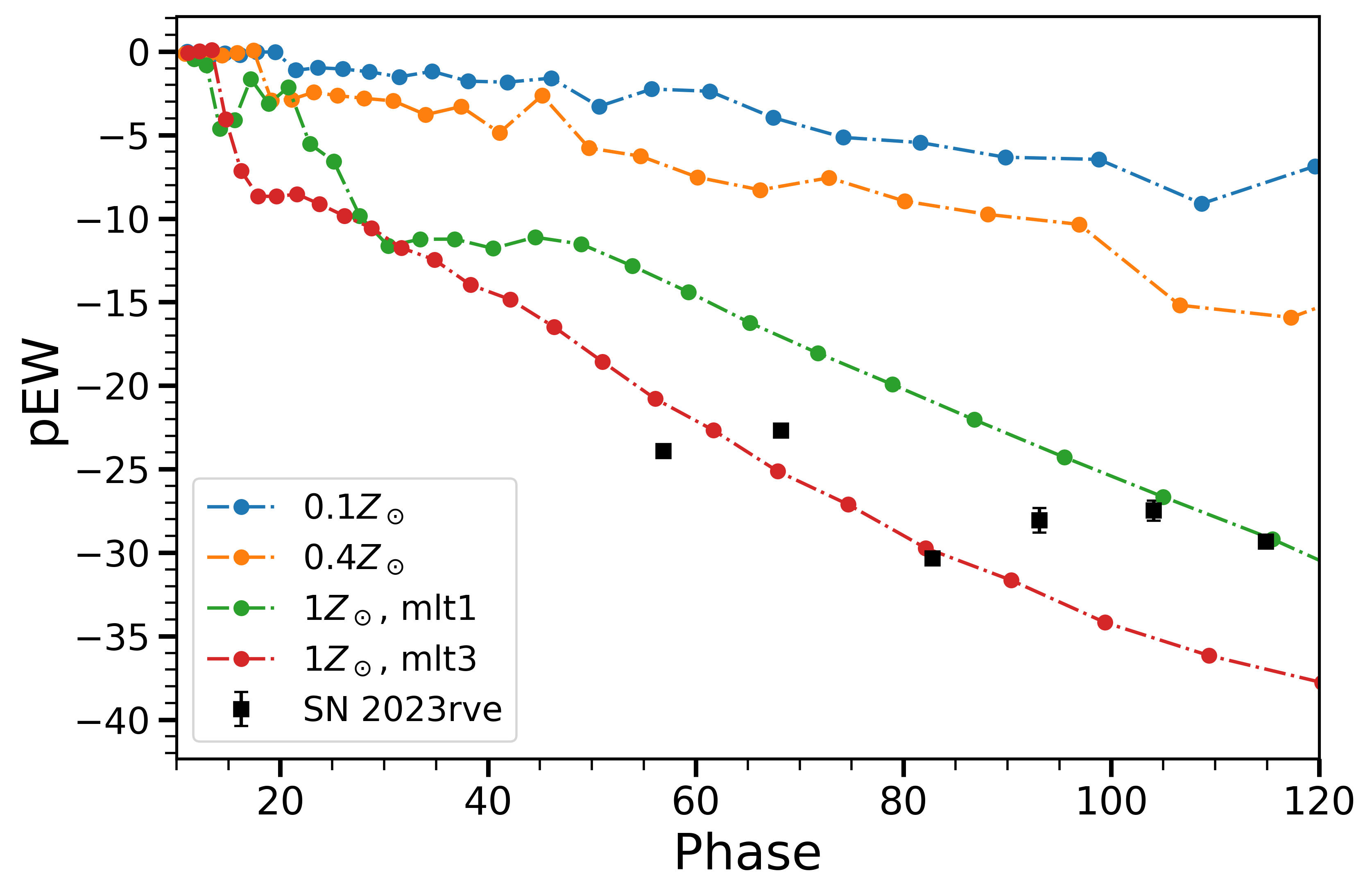}
\caption{SN\,2023rve \FeII{} line pseudo-equivalent widths alongside \cite{dessart2014} models of known metallicity. The models have 0.1 \(Z_\odot\), 0.4 \(Z_\odot\), and 1 \(Z_\odot\). The solar metallicity models have main sequence mass 15 \(M_\odot\) and two different mixing length parameters (mlt).
\label{fig:metal}}
\end{figure}

We compare these measurements to the 15 \msun main-sequence star models presented in \cite{dessart2014}, which span metallicities of 0.1, 0.4, and 1 \zsun. This comparison is shown in Figure \ref{fig:metal}. Two 1 \zsun models are shown with different mixing length parameters (mlt) in the stellar evolution stage, which control the efficiency of convective energy transport. Overall, the observed \FeII{} pEW evolution is most consistent with the approximately solar-metallicity models.

\section{Observational Properties}\label{sec:obs_properties}
\subsection{Distance}

Several measurements of the distance to NGC 1097 exist using the Tully-Fisher relationship \citep{tf1977}, which correlates the luminosity and rotational velocity of spiral galaxies. The most recent gives a distance modulus to NGC 1097 of $30.94 \pm 0.45$ mag, corresponding to $15.4 \pm$ 3.2 Mpc \citep{Tully2016}. 

Since individual galaxies can deviate from this relation \citep{czerny2018astronomical}, we also apply the expanding photosphere method \citep[EPM;][]{Kirshner1974}, following the methodology and implementation of \cite{Dong2021}. EPM solves for the angular size $\theta$ and color temperature $T_c$ from multiband photometry, which combined with the photospheric expansion velocity $v_\mathrm{phot}$ yields a distance $D$ and explosion epoch $t_0$ via

\begin{equation}
    t = D\left(\frac{\theta}{v_\mathrm{phot}}\right) + t_0,
\end{equation}

using the dilution factors of \cite{Dessart2005A&A...439..671D} and filter subsets $\{BV\}$, $\{BVI\}$, and $\{VI\}$. We note that these dilution factors were derived from atmosphere models that do not account for CSM interaction, which may introduce systematic uncertainties in the derived distance, particularly at early times when CSM influence is strongest. We also noticed a small difference in the slopes for the EPM data that may be related to the presence of CSM at early times. However, even after computing the distance for the two separate slopes and fixing the explosion epoch, the results are still consistent with the distance adopted in this paper. 



\begin{figure}
\includegraphics[width=1.\linewidth]{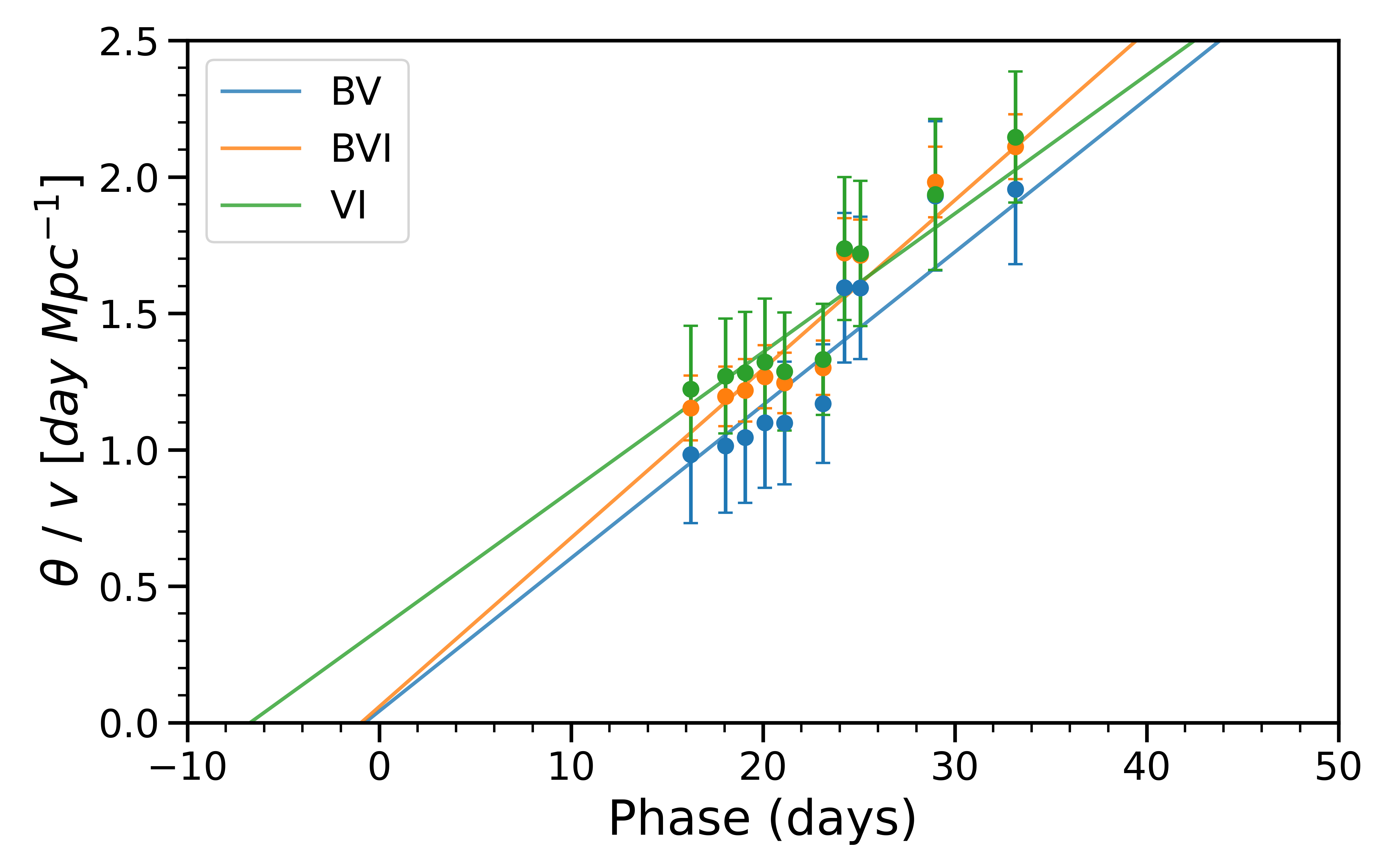}
\caption{EPM fit for SN\,2023rve using filter subsets \{BV\}, \{BVI\}, and \{VI\}. The resulting distances are 18.4 $\pm$ 7.0 Mpc, 16.3 $\pm$ 2.5 Mpc, and 19.7 $\pm$ 5.3 Mpc, respectively. 
\label{fig:epm}}
\end{figure}

The photospheric velocity used for the EPM can be estimated by measuring the minimum of the P-Cygni profile for \FeII{} $\lambda$5169 (Table \ref{tab:fevel}). While \FeII{} is more reliable than \Ha{} as a photospheric velocity tracer, \cite{Dessart2005A&A...439..671D} showed that it can still underestimate the true photospheric velocity by roughly 5--10\%, which therefore represents a systematic uncertainty in the EPM method. Since the necessary relation becomes nonlinear after around 40 days \citep{jones2009distance}, we restrict our analysis to the ten spectra between 16.24 and 33.16 days after explosion. 

Using photometric data collected on the same days and further interpolated photometric data using a Gaussian process, 3 filter subsets were analyzed: \{BV\}, \{BVI\}, and \{VI\}. The \textit{rp} and \textit{ip} magnitudes were converted to I magnitudes using the conversion equations in \cite{lupton2005}. Each of these filter subsets produced results for the distance to SN\,2023rve: 18.4 $\pm$ 7.0 Mpc, 16.3 $\pm$ 2.5 Mpc, and 19.7 $\pm$ 5.3 Mpc, respectively. This analysis is shown in Figure \ref{fig:epm}. Since these results are consistent with the most recent published values from the Tully-Fisher method, we adopt that distance, 15.4 $\pm$ 3.2 Mpc, for the remainder of this paper. 

\begin{deluxetable}{cc}
\tablecaption{Velocities of \FeII{} $\lambda$5169 used for EPM fitting
\label{tab:fevel}}
\tablewidth{0pt}
\tablehead{
\colhead{Date}&
\colhead{\FeII{} $\lambda$5169 Velocity (km s$^{-1}$) 
}
}
\startdata
2023-09-20 & 7080 $\pm$ 85 \\
2023-09-22 & 7080 $\pm$ 95 \\
2023-09-23 & 7080 $\pm$ 65 \\
2023-09-24 & 6906 $\pm$ 43 \\
2023-09-25 & 7138 $\pm$ 49 \\
2023-09-27 & 6964 $\pm$ 25 \\
2023-09-28 & 5339 $\pm$ 60 \\
2023-09-29 & 5397 $\pm$ 24 \\
2023-10-03 & 4759 $\pm$ 28 \\
2023-10-07 & 4643 $\pm$ 50
\enddata
\end{deluxetable}

\subsection{Explosion Epoch}
To constrain the explosion epoch of SN\,2023rve, we take the middle of the last nondetection available (ATLAS, JD 2460189.506, limiting magnitude of 18.39) and the first detection in prediscovery imaging (Telescope Live, JD 2460193.540, R$\approx$14.4 mag). The ATLAS limit is approximately 4 mag deeper than the source at first detection, indicating that the nondetection can be considered a robust constraint. This gives an explosion epoch of JD 2460192.0 $\pm$ 2.0. The EPM can also be used to constrain the explosion epoch by fitting a value for $t_0$ in each filter subset. We obtain values from the \{BV\}, \{BVI\}, and \{VI\} subsets of JD 2460190.666 $\pm$ 7.66, 2460190.849 $\pm$ 3.26, and 2460185.380 $\pm$ 7.22, respectively. Due to the size of the error bars on these results, we conclude that the explosion epoch derived from the EPM is consistent with what was derived from the last nondetection and prediscovery imaging, and we maintain this epoch for the remainder of this paper.  

\section{Photometric Evolution}\label{sec:phot_evol}

The multiband photometric evolution of SN\,2023rve is shown in Figure \ref{fig:lightcurve}. After first detection, the V band magnitude brightens slightly before reaching a maximum apparent magnitude of 14.30 mag (absolute magnitude of -17.11 $\pm 0.46$ mag) on JD 2460200.071, approximately 8 days after explosion. 
Following this maximum, recombination of the hydrogen envelope causes a slow decline which continues until around day 140 (plateau phase).
At this point the light curve exhibits a steep drop-off and then continues to steadily decline until no longer visible. Other bands follow these same patterns closely. 

A noticeable bump-like feature is present in the optical light curves during the transition from the plateau to the radioactive tail phase, at epochs of approximately 140--160 days after explosion. Since this feature is observed consistently in multiple photometric bands, it is unlikely to be a photometric artifact. Similar post-plateau luminosity excesses have been reported in a small number of Type~II supernovae, most notably SN\,2005cs and SN\,2009bw. Such deviations from a smooth plateau-tail transition have been interpreted as the result of the complex interplay between the receding hydrogen recombination front and radioactive heating from $^{56}$Ni and its decay products. In particular, the morphology of the transition is expected to be sensitive to the degree of $^{56}$Ni mixing and to the density and compositional structure of the inner ejecta. The temporal coincidence of the observed bump with the onset of the radioactive tail suggests that these effects may contribute to the observed luminosity excess. Alternative scenarios, including weak circumstellar interaction or fallback accretion, cannot be completely excluded. However, the available photometric observations do not provide compelling evidence for an additional energy source, and therefore an interpretation in terms of ejecta structure and radioactive-energy deposition appears to be the most plausible explanation.

\begin{figure}
\includegraphics[width=1.\linewidth]{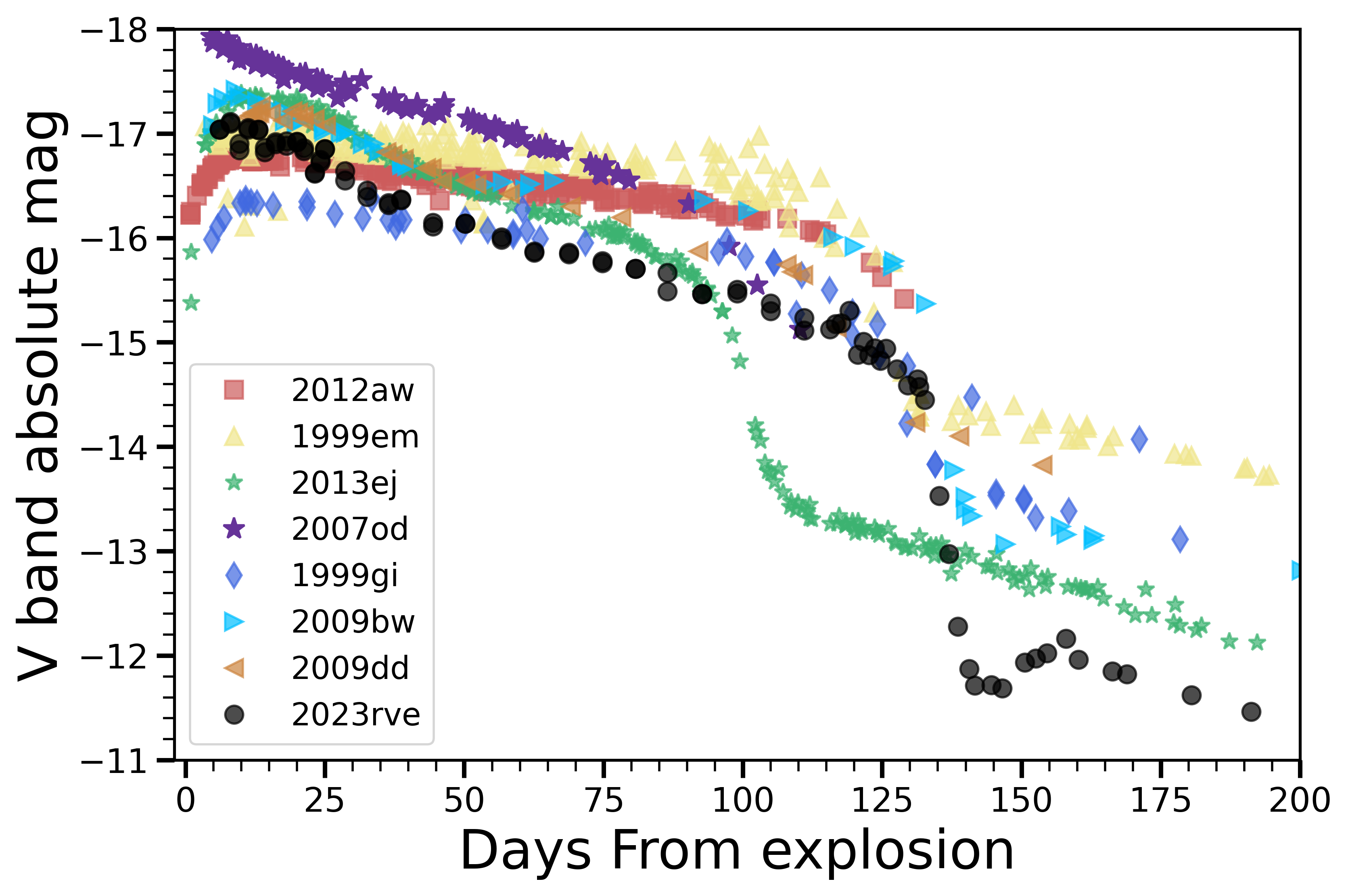}
\caption{V-band light curves of SN\,2023rve and other similar Type II supernovae. 
\label{fig:compares}}
\end{figure}

\begin{figure}
\includegraphics[width=1.\linewidth]{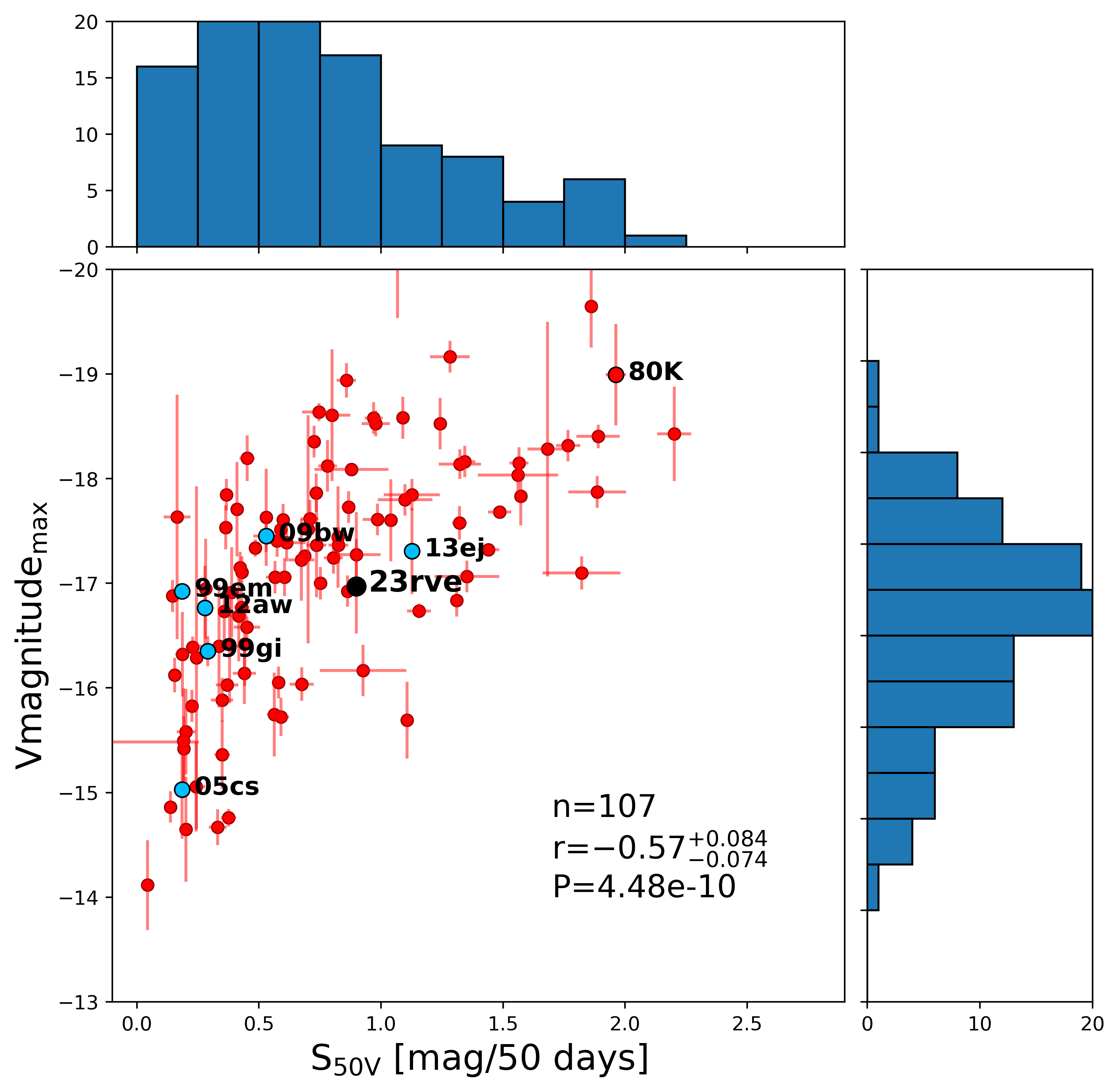}
\caption{The $M_V$ compared to $S_{50V}$ for SN\,2023rve and a sample of SNe II \citep{anderson2014characterizing, Valenti2016} from SNDAVIS. SN\,2023rve is shown in black, and is near the middle of the SNe IIP and SNe IIL range. The sample referenced in this paper is shown in blue, other notable objects are in green, and the rest of the sample is in red.
\label{fig:compare}}
\end{figure}

\begin{figure}
\includegraphics[width=1.\linewidth]{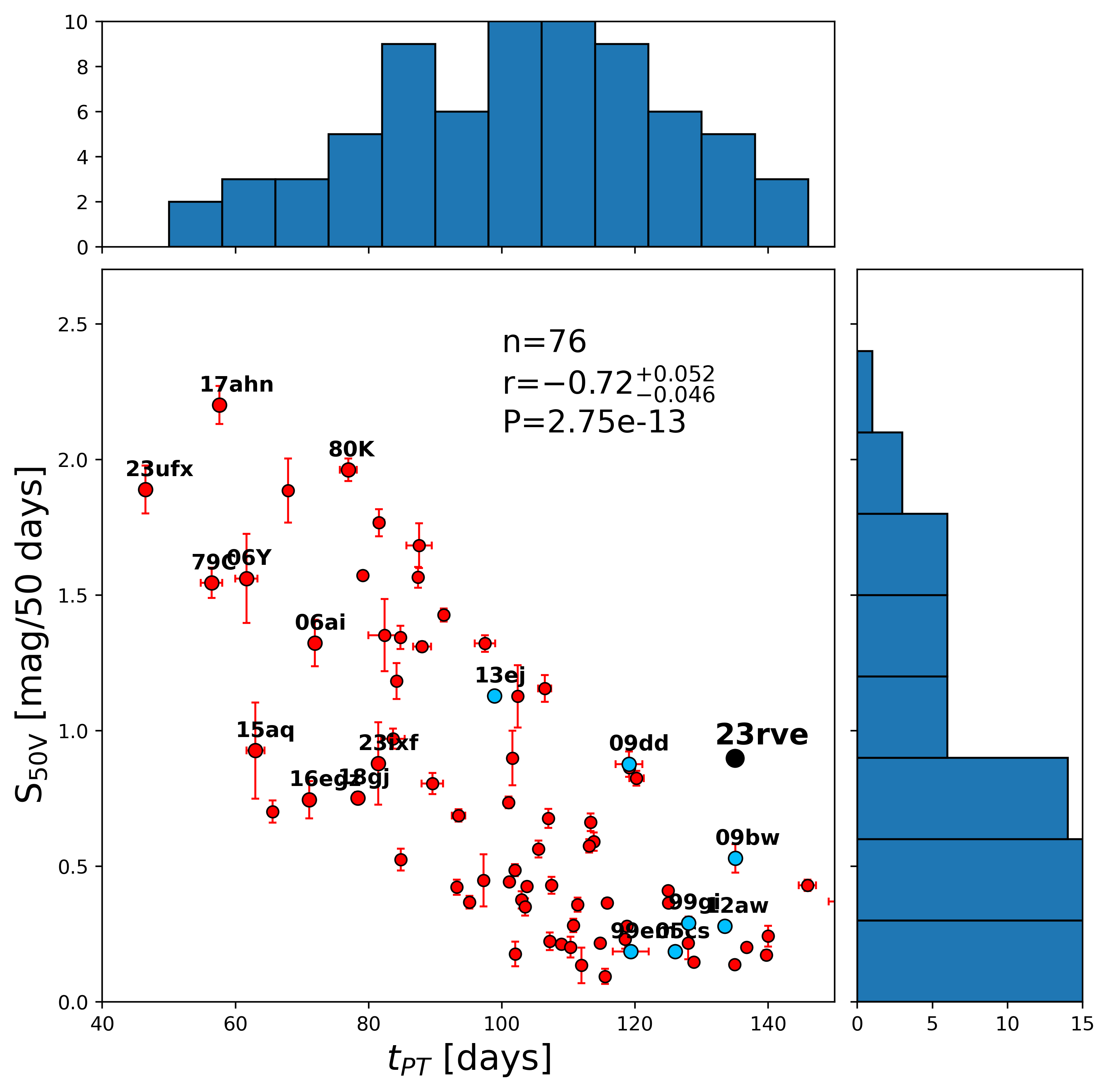}
\caption{The plateau length, t$_{PT}$ versus the slope S$_{50V}$ for SN\,2023rve and a sample of SNe II.
\label{fig:tpt}}
\end{figure}

This V-band light curve is shown in Figure \ref{fig:compares}, along with the V-band light curves of similar objects for comparison. The length of the plateau in SN\,2023rve is similar to that of SN\,1999em, SN\,2012aw, and SN\,2009bw, and the slope is similar to that of SN\,2007od or a nearly Type IIL SN, like SN\,2013ej. 
\\
In order to classify Type II SNe along the spectrum of SNe IIP-like and SNe IIL-like objects, we can compare statistical properties, such as the decline rate after maximum brightness, with other SNe II. Using \cite{Valenti2016}, the V-band slope per 50 days ($S_{50V}$) of SN\,2023rve was measured to be 0.899 $\pm$ 0.015 $\rm mag\,(50\,days)^{-1}$. \cite{anderson2014characterizing} suggests a strong correlation between the maximum absolute magnitude in the V-band and the decline rate ($S_{50V}$). Figure \ref{fig:compare} shows this relation for a sample of SNe II, including SN\,2023rve. In this plot, SNe IIP-like objects are typically farther to the left with longer, flatter plateaus, while SNe IIL-like objects are farther to the right with faster decline. SN\,2023rve falls roughly in the middle, and almost appears to be Type IIL-like given its rapid decline rate. However, the length of the plateau is rather long for Type IIL-like SNe ($t_{PT}$ = 135.03 $\pm$ 0.27 days)\footnote{We use the $t_{PT}$ definition from \cite{Valenti2016}}. In this regard, SN\,2023rve is similar to SN\,2009bw, which also has a long plateau with a Type IIL-like slope (see Figure \ref{fig:tpt}).

We also compare the value $a_0$, the depth of the drop from plateau into the tail phase, with the sample of SNe II.
SN\,2023rve has a relatively high $a_0$ compared to the sample, but not as high as SN\,2013bu, ASASSN-14ha, and SN\,2005cs, which are all low-luminosity objects.

After the fall from plateau, the SN enters the radioactive tail phase, powered at this point primarily by $^{56}$Co decay. If the gamma rays from the decay are fully trapped in the ejecta, the luminosity declines at a rate of about 1 mag per 100 days \citep{1988ApJ...330..218W}; however, incomplete trapping can lead to a faster decline. For SN\,2023rve, the radioactive tail decline is consistent with nearly complete gamma-ray trapping, with only a mild deviation from the complete trapping rate. In Section \ref{sec:nickel}, we describe the construction of the pseudo-bolometric light curve and the method used to derive an estimate of the nickel mass from the radioactive tail.

\begin{figure}
\includegraphics[width=1.\linewidth]{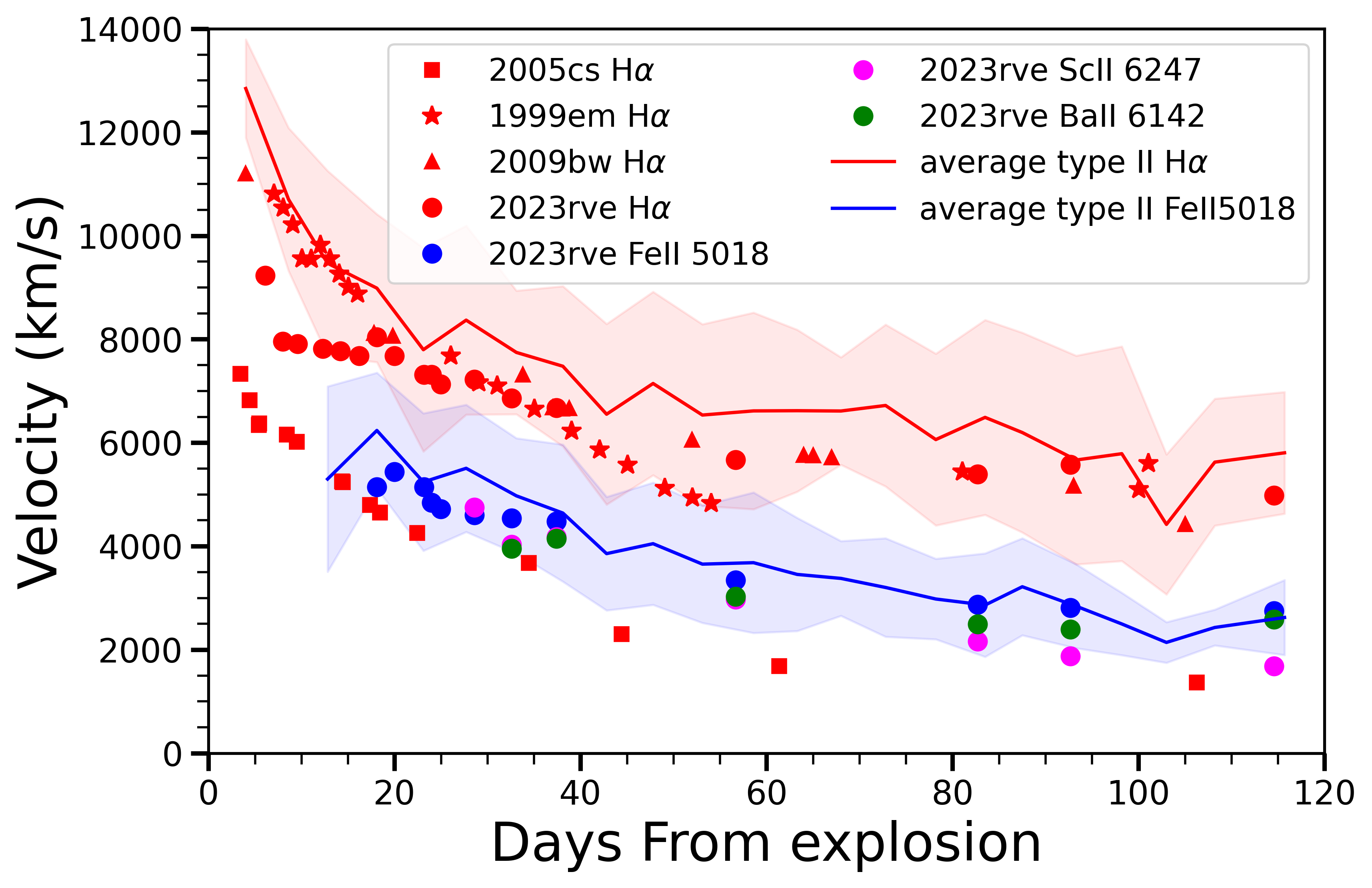}
\caption{Expansion velocity evolution of \Ha{}, \FeII{}, \ScII{}, and \BaII{} from SN\,2023rve and \Ha{} velocities for other objects from the sample alongside the mean of a sample of SNe II \Ha{} and \FeII{} $\lambda$5018 velocities \citep{gutierrez2017type}. One standard deviation is shaded. 
\label{fig:velocity}}
\end{figure}


\section{Spectroscopic Evolution}\label{sec:spec}
\subsection{Optical Spectra}\label{optic}
The optical spectra of SN\,2023rve are shown in Figure \ref{fig:evolution}. At early times, a blue continuum dominates with a broad, shallow \Ha{} absorption feature. Around day 14, \FeII{} $\lambda$5018 emerges and can be used to trace the photospheric velocity, consistent with the cooling and expansion of the photosphere. Around day 30, additional metal lines appear including \ScII{} ($\lambda$6247) and \BaII{} ($\lambda$6142), which are commonly seen in SNe II \citep{gutierrez2017type}, as the photosphere cools sufficiently for these lower-ionization species to form. The spectra continue to become redder and the hydrogen lines develop P Cygni features, and at later times the spectra become emission line-dominated as the SN enters the nebular phase.

The evolution of the expansion velocity for these four important tracers in the SN's ejecta (\Ha{} $\lambda6563$, \FeII{} $\lambda$5018, \ScII{} $\lambda$6247, and \BaII{} $\lambda$6142) are shown in Figure \ref{fig:velocity}. Line velocities were measured by selecting spectral regions around the feature of interest, normalizing and smoothing the flux, and then fitting a low-order polynomial to the P-Cygni absorption minimum. The minimum of this fit was used to determine the observed wavelength of the line, from which the velocity shift relative to the rest wavelength was calculated. To isolate the H$\alpha$ emission from the SN (versus that of the host galaxy and the \NII{} line), we fit a triple Gaussian to the nebular spectra. For the nebular spectra utilized in Figure \ref{fig:velocity}, the standard deviation from a Gaussian fit for an emission line can be converted to the full width at half maximum (FWHM) for this purpose through the equation 
\begin{equation}
    \mathrm{FWHM} = 2\sqrt{2\ln{2}}\sigma\approx2.3548\sigma
\end{equation}
SN\,2023rve shows relatively low-velocity \Ha{}, particularly at early times where the signature is more broad and shallow. The velocity falls within one standard deviation of the mean for a larger sample of SNe II \citep{gutierrez2017type} for the remainder of its evolution. The \FeII{} velocity of SN\,2023rve is more consistent with the average from the sample. 

Figure \ref{fig:velocity} shows a comparison of the evolution of the ejecta velocity of SN\,2023rve alongside similar objects and the mean of a larger sample of SNe II \citep{gutierrez2017type}, using the different techniques outlined above. This comparison reveals that SN\,2023rve has a low velocity relative to other SNe II in the early and plateau phases. Notably, during the nebular phase (not included in Figure \ref{fig:velocity}) the \Ha{} ejecta velocity of SN\,2023rve (220 km s$^{-1}$) is even lower than the low-luminosity, low-velocity SN\,2005cs (730 km s$^{-1}$) at a similar epoch. This raises the question of whether SN\,2023rve belongs in the low-luminosity class of SNe II. The long plateau and low nickel mass may also point in this direction, while the plateau slope may argue against this low-luminosity interpretations. To investigate this possibility we compared SN\,2023rve to the low-luminosity sample presented in \cite{spiro2014low}. 

\begin{figure}
\includegraphics[width=1.\linewidth]{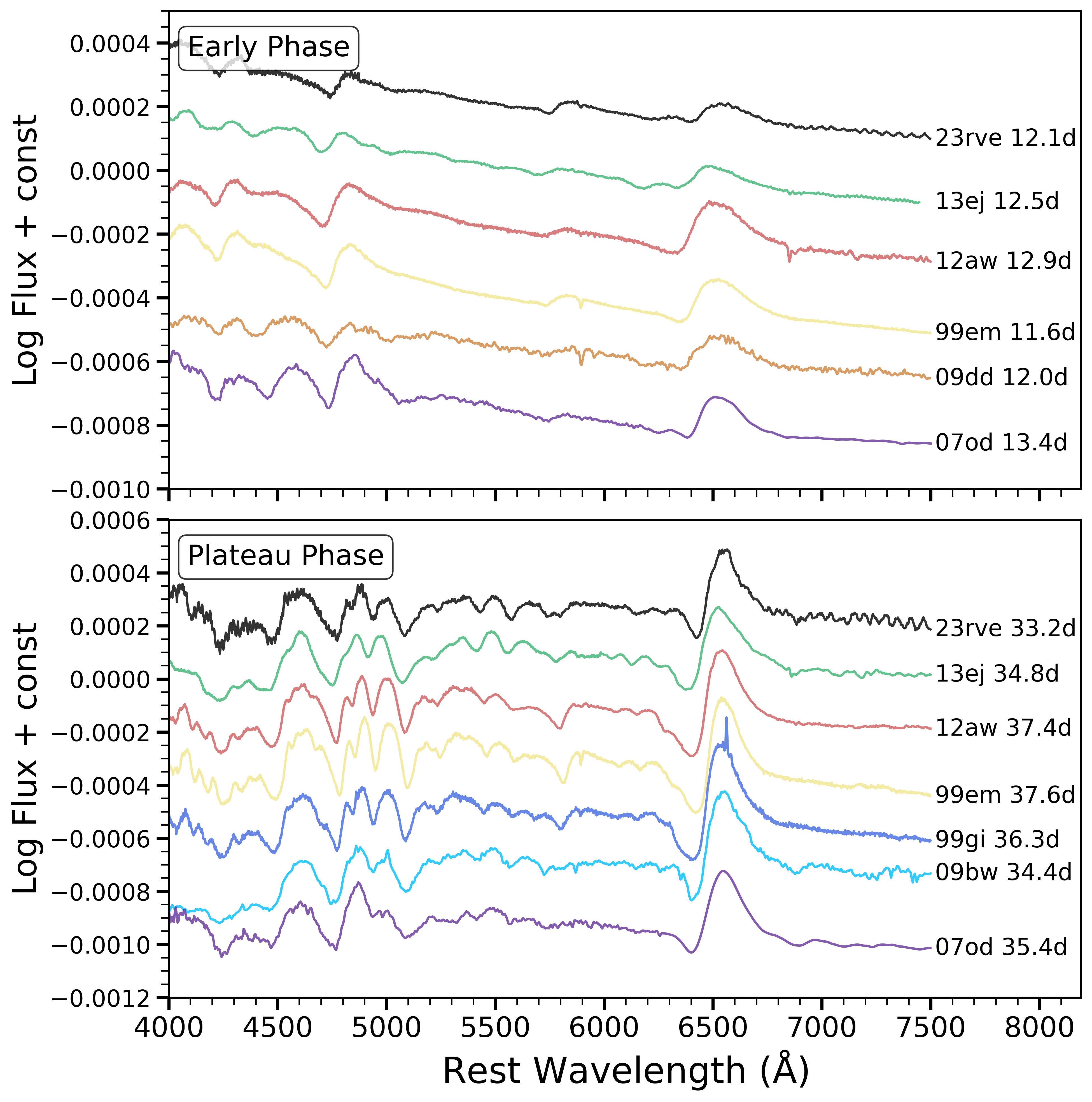}
\caption{Spectral comparisons during the early (top) and plateau (bottom) phase between SN\,2023rve and similar objects. The \Ha{} absorption minimum of SN\,2023rve is less blueshifted than most comparison objects at both epochs, consistent with the lower expansion velocities shown in Figure \ref{fig:velocity}.
\label{fig:speccomp_early}}
\end{figure}

\begin{figure}
\includegraphics[width=1.\linewidth]{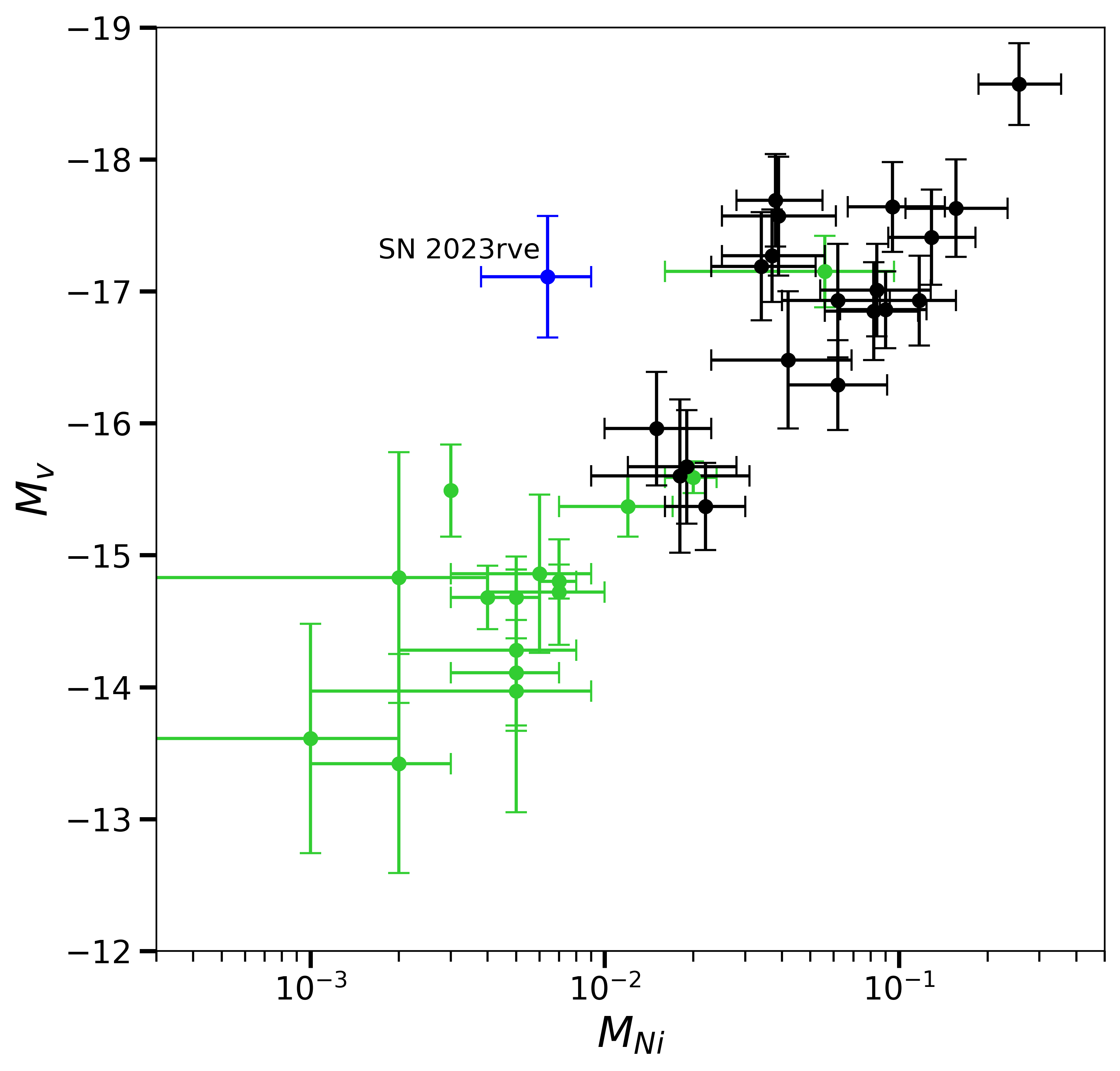}
\caption{The peak V-band magnitude $M_V$ and $^{56}$Ni mass $M_{Ni}$ for SN\,2023rve (blue) relative to a sample of low-luminosity SNe IIP (green) and normal SNe IIP \citep{spiro2014low}. SN\,2023rve is an outlier with a peak luminosity that is bright for its low nickel mass.
\label{fig:mv_mni}}
\end{figure}

\begin{figure}
\includegraphics[width=1.\linewidth]{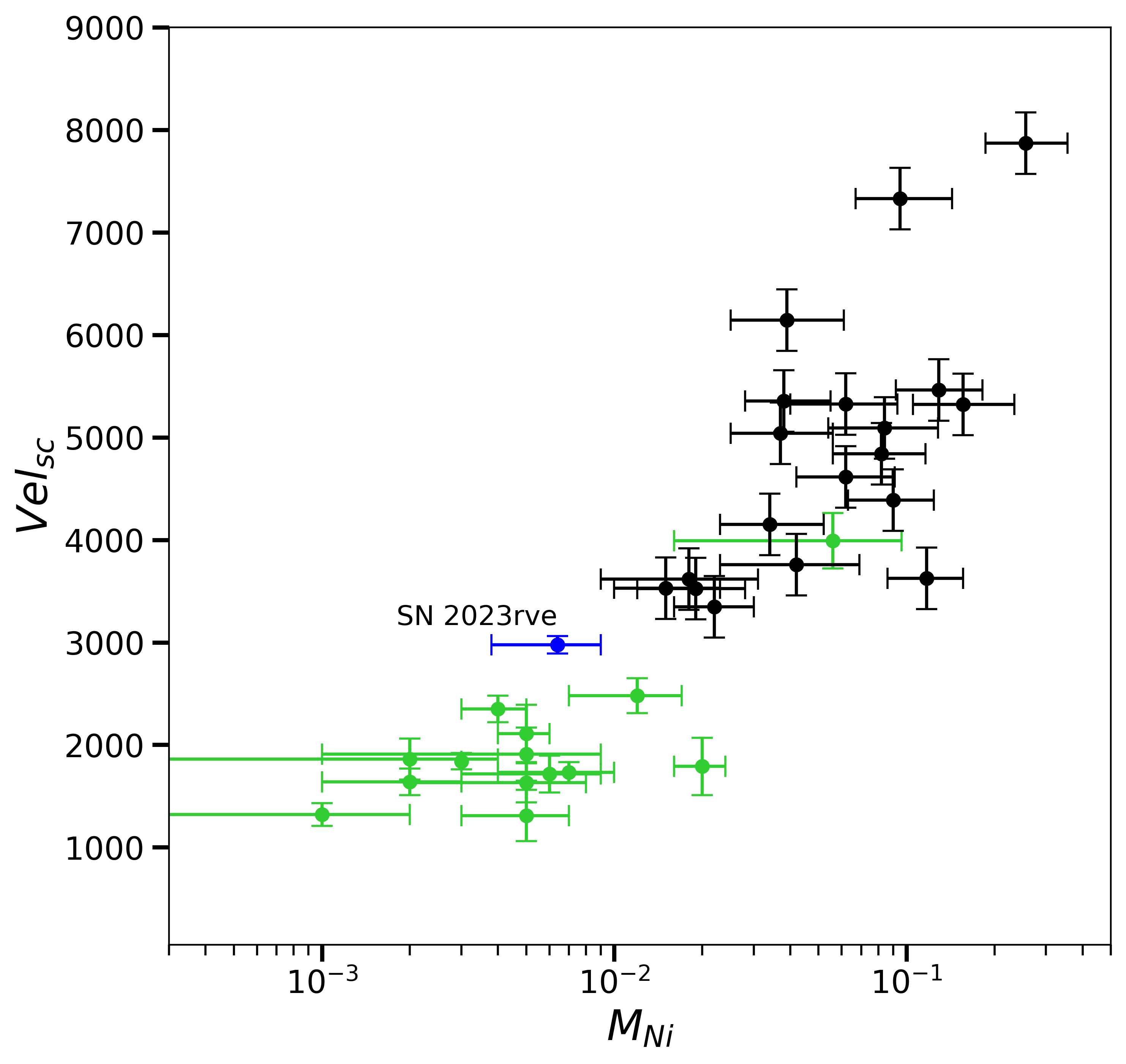}
\caption{\ScII{} velocity versus $^{56}$Ni mass for the same sample as Figure \ref{fig:mv_mni}. SN\,2023rve falls broadly within the overall trend, but lies closer to the low-luminosity sample, with a velocity and nickel mass lower than most normal SNe IIP.
\label{fig:mv_velsc}}
\end{figure}

Figures \ref{fig:mv_mni} and \ref{fig:mv_velsc} compare the peak absolute magnitude and \ScII{} velocity, respectively, against nickel mass for a sample of normal and low-luminosity SNe IIP \citep{spiro2014low}. SN\,2023rve is an outlier in the luminosity-nickel relationship, with a peak luminosity that is bright relative to its nickel mass, while in the velocity-nickel plane it falls broadly within the overall trend but closer to the low-luminosity sample. This further supports the picture of SN\,2023rve combining characteristics of both normal and low-luminosity SNe IIP. The early and plateau phase spectra of SN\,2023rve are compared to similar SNe II in Figure \ref{fig:speccomp_early} to illustrate the spectroscopic comparison discussed above. 

\begin{figure}
\includegraphics[width=1.\linewidth]{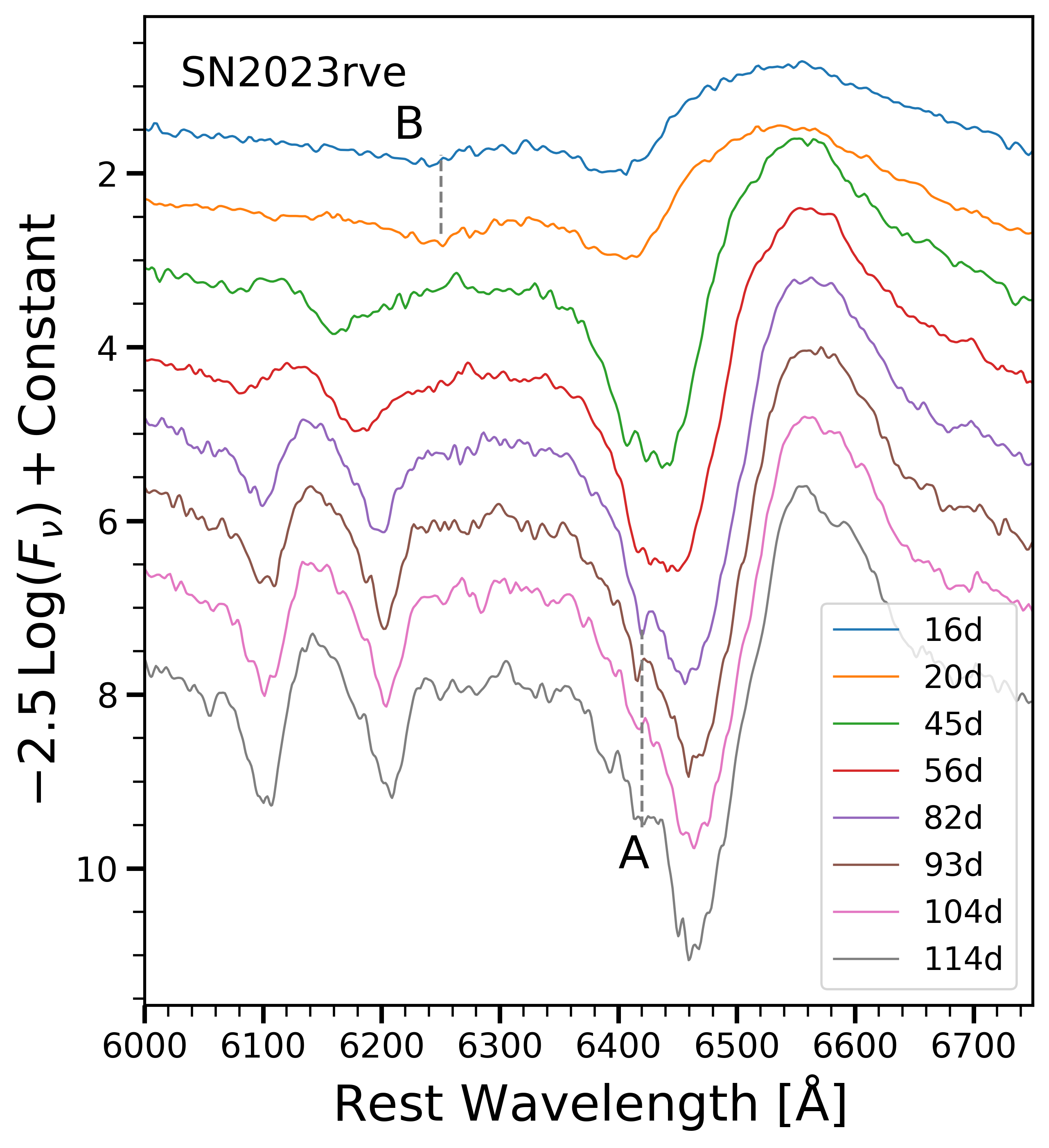}
\caption{The Cachito feature labeled as it appears at early time (likely \SiII{} $\lambda$6355), disappears in the intermediate stages, and reappears closer to \Ha{} (likely high-velocity \Ha{} from CSM interactions with ejecta). 
\label{fig:cachspace}}
\end{figure}

We observe the evolution of an extra absorption feature on the blue side of the \Ha{} signature known as Cachito \citep{gutierrez2017type}, which appears in the first 20 days around $\lambda$6247 (denoted B in Figure \ref{fig:cachspace}). In this phase it is suspected to be \SiII{} $\lambda6355$ \citep{pastorello2006sn} due to its velocity being comparable to other metal velocities. The feature disappears and then reappears closer to the \Ha{} P Cygni profile around day 35 at $\lambda$6300 (denoted A in Figure \ref{fig:cachspace}), which could be attributed to high-velocity hydrogen because its velocity is comparable to the \Ha{} velocity at early times \citep{gutierrez2017type}. With this understanding of the feature's origins, the velocity of Cachito at early times (\SiII{} $\lambda$6355) is found to be $\sim$5000 km s$^{-1}$. At later times, Cachito is explained by CSM interaction with the SN ejecta, creating a dense shell with high-velocity \Ha{} emissions. The velocity of this later feature is found to be between $\sim$10000-11000 km s$^{-1}$. Clear evidence of high velocity hydrogen in \Hb{} is not visible. However, the \Hb{} region is quite crowded and the line is relatively weak compared to \Ha{}, making the detection of a high velocity feature more complicated.

\subsection{Nebular Spectra}\label{sec:nebular}

\begin{figure*}
\centering
\includegraphics[width = 1 \linewidth]{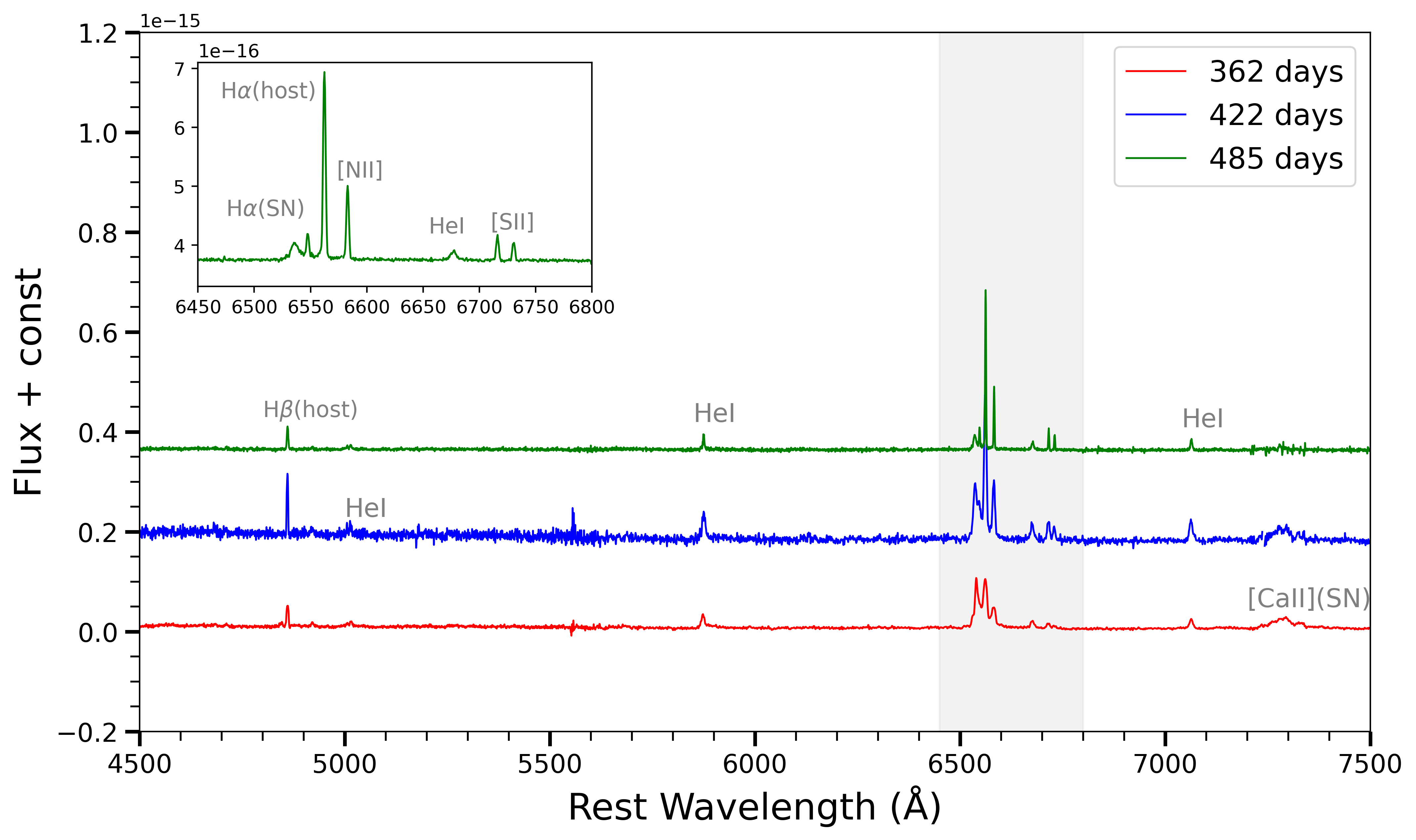}
\caption{Spectroscopic evolution of SN\,2023rve in the nebular phase, with identifiable emission lines labeled. The inset zooms in on the region of interest to show the resolution of the \HeI{} lines and the distinction of \Ha{} from the host and SN.
\label{fig:nebs}}
\end{figure*}

\begin{figure}
\includegraphics[width=1.\linewidth]{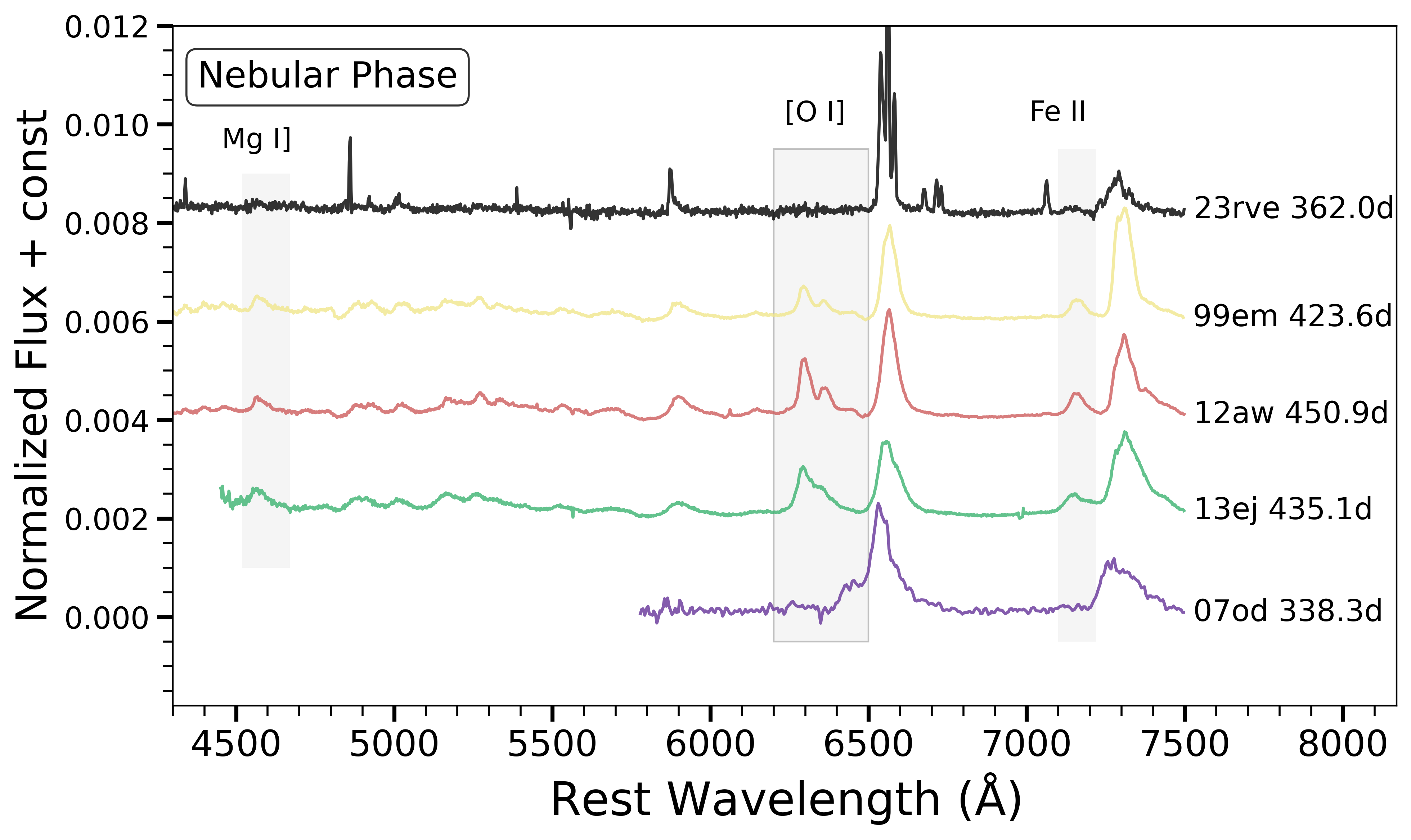}
\caption{Spectral comparison during the nebular phase between SN\,2023rve and similar objects. It is clear that, although at late time the helium is resolved, the oxygen signature is still not visible, and the velocity becomes notably lower. \FeII{} and \MgI{}] are also seen in the nebular spectra of similar objects but appear very weakly for SN\,2023rve.
\label{fig:speccomp_late}}
\end{figure}

We obtain 3 spectra from the nebular phase, after the fall from plateau, at days 362, 422, and 485 after explosion. These nebular spectra are shown in Figure \ref{fig:nebs}, and compared with the sample in Figure \ref{fig:speccomp_late}. The nebular spectra of SN\,2023rve are characterized by a relatively small number of emission features and unusually narrow line profiles. The spectra are dominated by \Ha{} and \CaII{} emission, while several lines commonly observed in nebular SNe II, including [\OI{}], \MgI{}], and [\FeII{}], are weak or absent. In addition, narrow \HeI{} emission is detected throughout the nebular phase.

A notable feature of the nebular spectra is the relatively flat continuum baseline. This raises the possibility that part of the apparent continuum originates from residual host-galaxy contamination or noise rather than true SN emission. To investigate this possibility, we repeated the photometric reduction using template-subtracted imaging in order to eliminate any residual host-galaxy contamination. We also  extracted a spectrum of the host galaxy near the SN position and subtracted the host spectrum from the nebular spectra.
The template-subtracted photometry (e.g. at 362 days, $B\sim22.2, V\sim22$ compared to $B\sim21.9, V\sim21.5$ without template subtraction) is most consistent with a host-subtracted nebular spectrum. However, the shape and intensity of the nebular emission lines do not change significantly (after subtracting the host spectrum) confirming that [\OI{}], \MgI{}], and [\FeII{}], are weak or absent and the flat continuum is most likely an intrinsic characteristic of the SN and not residual host-galaxy contamination.

No forbidden [\OI{}] $\lambda 6300, \lambda 6364$ is visible in the nebular spectra collected. The absence of these lines is intriguing, since oxygen is one of the most abundant elements in SNe, and is usually clearly detected in SNe II spectra during late stages. Forbidden lines such as [\OI{}] require sufficiently low electron densities to emit, as collisional de-excitation suppresses them at high densities. One possibility is therefore that the ejecta had not yet reached the low-density nebular regime by day 362. However, this is difficult to reconcile with the simultaneous detection of other nebular indicators such as \CaII{} and \HeI{}. 

\begin{figure*}
\includegraphics[width=1.\linewidth]{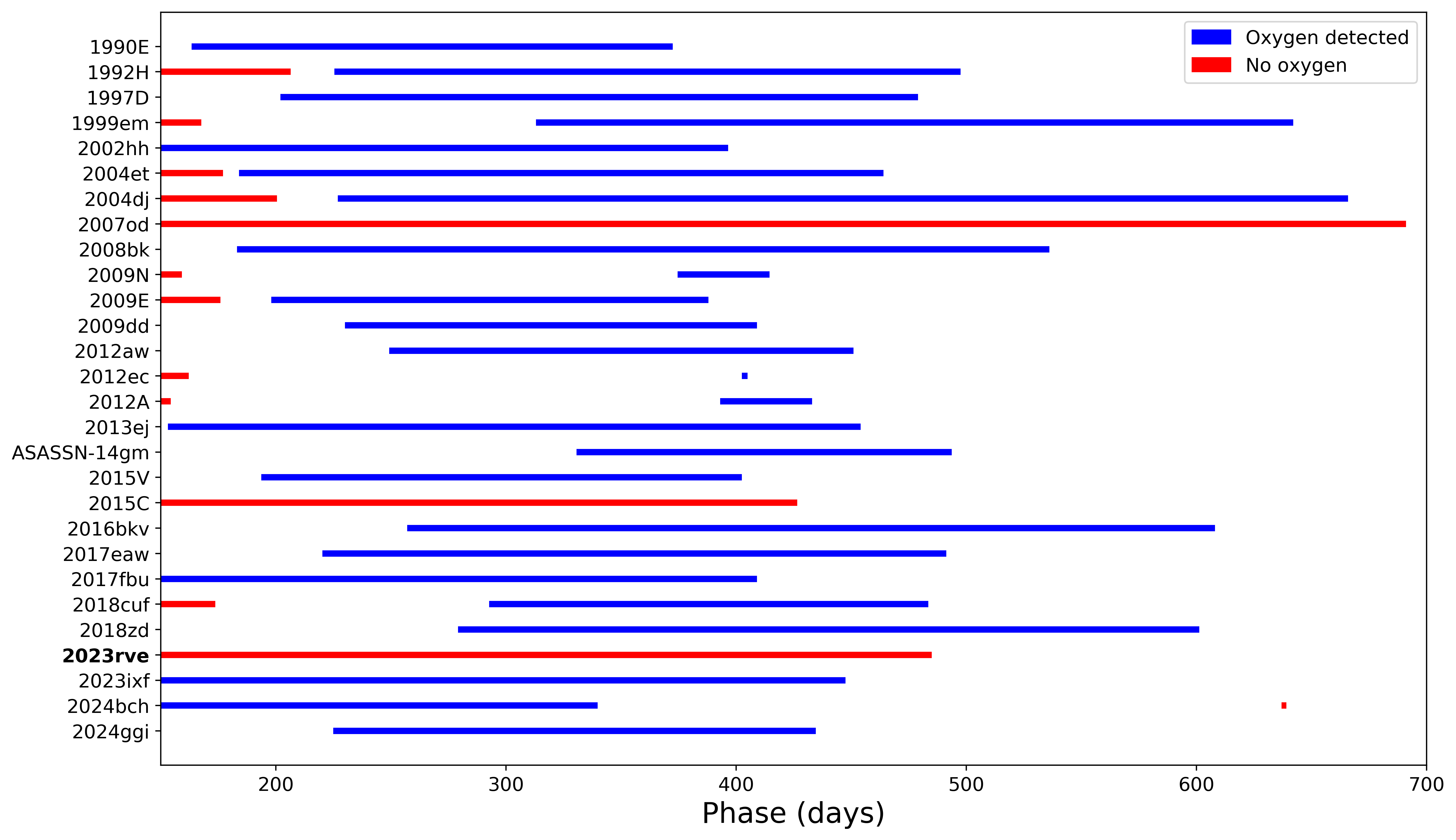}
\caption{Duration of [\OI{}] detections in nebular spectra between 150 and 700 days for Type II and Type II-like objects in the database. Objects with more than 3 spectra over a wide range of days, with a wavelength range including the [\OI{}] $\lambda6300,6364$ range were used in this sample. Only 2 other events (SN\,2007od and SN\,2015C) are also known to not show oxygen at these phases.
\label{fig:oIItimeline}}
\end{figure*}

We collected 2 more nebular spectra at 422 and 485 days after explosion and, while the \CaII{} and \Ha{} get weaker with time, there is still no sign of the [\OI{}], while helium lines and residual lines from the host are still detected. To confirm the uniqueness of this lack of oxygen, we checked for the presence of oxygen lines in nebular spectra of published data between day 140 and 700 for Type II SNe. We included in this sample only objects with nebular spectra between 362 and 485 days after explosion (the range available for SN\,2023rve). Only two objects (SN\,2015C and SN\,2007od) out of the 26 SNe we looked at are also seemingly missing [\OI{}] in the available time window (140-700 days after explosion; see Figure \ref{fig:oIItimeline}). 

To the best of our knowledge, SN\,2023rve and these two other objects are the only SNe II that do not show any signs of oxygen lines in their nebular spectra. SN\,2015C has a higher velocity in both the photospheric and nebular phase, along with a stronger \Ha{} flux, which could be due to a more massive hydrogen envelope. We were unable to do a more detailed study of SN\,2015C due to poor data coverage. SN\,2007od is also a higher-velocity object than SN\,2023rve, with a bright peak magnitude of $M_V = -18$ mag, a short plateau, and a low ejected $^{56}$Ni mass \citep{inserra2011}. Evidence for new dust formation and CSM interaction are observed in its nebular spectra, with up to $\sim4\times10^{-4}\ M_\odot$ of dust implied by radiative transfer modeling \citep{2007odandrews}.

An alternative explanation of the lack of [\OI{}] could be CO molecular lines or dust formation. In both of these processes oxygen lines are suppressed \citep{Dessart_2025}. Dust forming in the dense inner ejecta can block emission from oxygen and calcium, while H$\alpha$ remains visible, as observed for example in SN\,2023ixf which formed dust at early times \citep{bostroem2025,jacobsongalan2023,singh2026}. This makes the SN appear like it ejected very little O or Ni, even if the elements are actually present.

Interestingly, both SN\,2023rve and SN\,2007od seem to show evidence of dust formation in their ejecta \citep{2007odandrews}. We also cannot exclude dust formation in SN\,2015C due to poor data coverage, and the presence of dust in two of the three objects suggests this may be a common factor in SNe II lacking nebular oxygen emission. 

Another possibility is that, as shown by \cite{Dessart2021}, if oxygen is well mixed with calcium in the inner ejecta, calcium becomes the dominant coolant and reduces the [\OI{}] $\lambda\lambda$6300, 6364 flux. The presence of [\CaII{}] $\lambda\lambda$7291, 7323 emission in the nebular spectra of SN\,2023rve is consistent with this scenario, though detailed nebular modeling would be required to determine whether the observed calcium emission strength is compatible with this explanation.

An alternative interpretation, suggested by the very weak \FeII{} and \MgI{} lines alongside the absent [\OI{}] emission, is that the true $^{56}$Ni mass may be even lower than our light curve estimate, or even zero. In the case of zero $^{56}$Ni, the regular thermal emission lines would be absent, and any observed emission could instead be powered by ongoing CSM interaction, which would preferentially excite the outer hydrogen and helium layers rather than the inner oxygen and calcium-rich ejecta 
\citep{1994ApJ...420..268C,2023A&A...675A..33D}. This would naturally explain the detection of \Ha{} and \HeI{} while other lines remain almost undetected. 

This interpretation could also be consistent with the fallback of oxygen-rich material onto a central compact object. In this scenario, some fraction of the inner ejecta fails to reach escape velocity and falls back onto the newly formed remnant, reducing the amount of oxygen-rich material contributing to nebular emission. Fallback explosions are also expected to produce relatively low explosion energies, low $^{56}$Ni masses, and weak nebular metal lines, broadly consistent with several observed properties of SN\,2023rve. However, if the light-curve is sustained primarily by CSM interaction rather than $^{56}$Ni decay, this would require further explanation for why the slope of the tail is nearly consistent with $^{56}$Ni--$^{56}$Co decay.

\begin{figure}
\includegraphics[width=1.\linewidth]{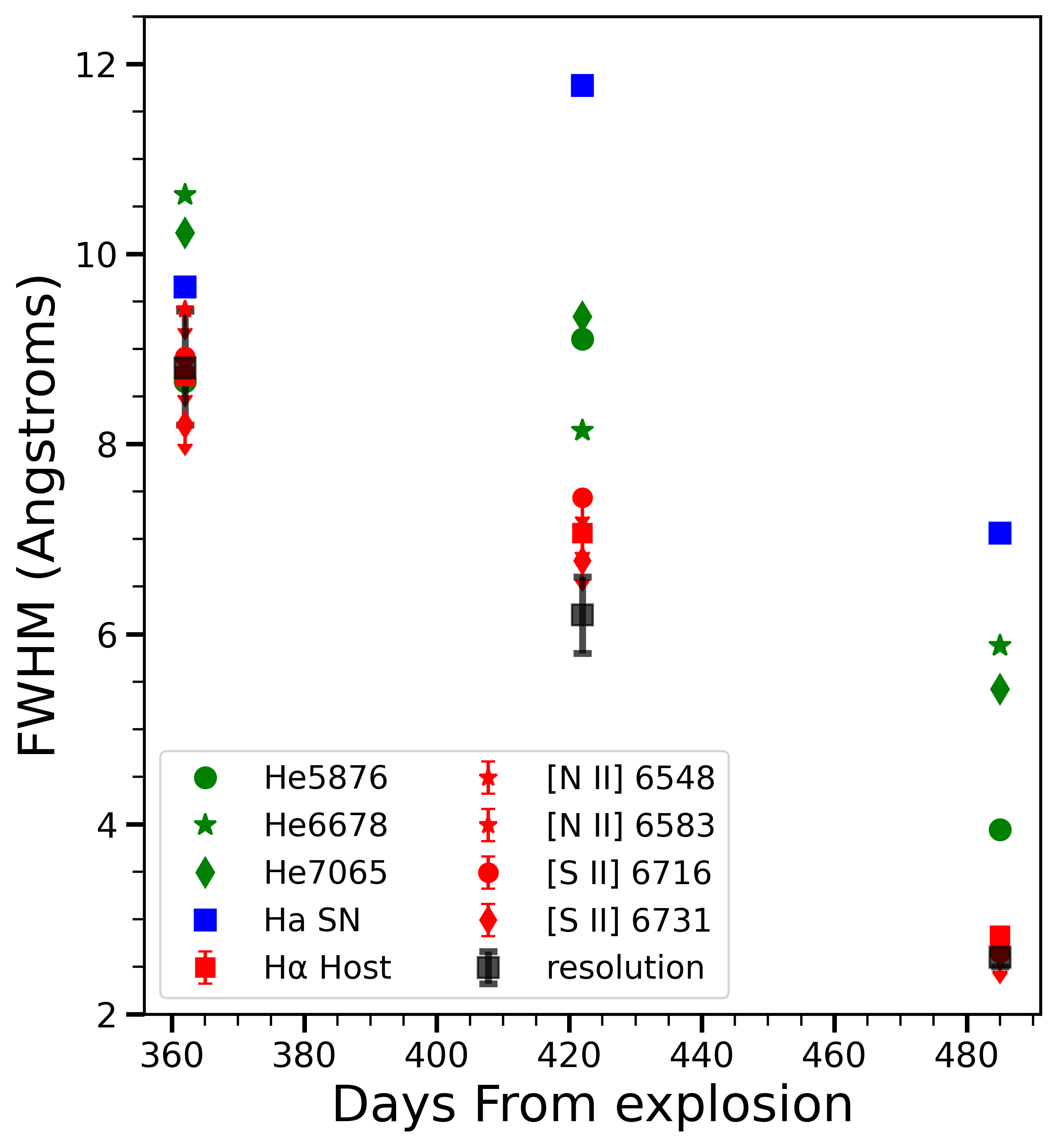}
\caption{Measured line widths for different spectral signatures in the early, plateau, and nebular phase. The resolution of each spectrum is shown in black, the red points refer to lines of host galaxy origin, the blue is from the SN, and the green (\HeI{}) is of unclear origin.
\label{fig:fwhm}}
\end{figure}

Helium lines are also not typical in the nebular spectra of SNe II, and in order to confirm their origin from the SN region, we used a higher-resolution setup for the spectrum at 485 days. The average spectral resolution was measured from sky lines in the regions of the \HeI{} features of interest, giving values of 8.8 $\pm$ 0.6, 6.2 $\pm$ 0.4, and 2.6 $\pm$ 0.1 \AA{} for the spectra at 362, 422, and 485 days respectively. FWHM values were measured for both SN and host galaxy features and are shown alongside the resolution in Figure \ref{fig:fwhm}.

The \HeI{} lines at $\lambda5876, 6678$, and 7065 exhibit FWHM values of 3.94, 5.87, and 5.42 \AA, respectively, while \Ha{} associated with the SN ejecta shows a FWHM of 7.06 \AA. In contrast, narrow host emission lines have smaller widths, with [\NII{}] and [\SII{}] lines clustered around 2.6 to 2.8 \AA. The proximity of these host values to the resolution of the spectra indicates they are unresolved. The higher instrumental resolution also explains the large change from earlier epochs, when SN and host features had FWHM values of 8 to 12 \AA. The FWHM of the \HeI{} lines falls between the FWHM of the SN and the host lines. While this would be consistent with being formed in the SN ejecta, the fact that their intensity does not decrease with time (like the \CaII{} and \Ha{} lines) may still support their formation in the SN environment. 

There are galaxies with helium in their emission spectra, but those are known to host AGN, which are the sources of the lines \citep{Mullaney_2008}. However, in the case of SN\,2023rve, the SN is not at the center of the galaxy, and there is no sign of an AGN. 

We also detect the forbidden [\CaII{}] $\lambda\lambda 7291, 7323$ doublet alongside the \CaII{} near-infrared triplet. Since the ratio of these two features is sensitive to the density of the calcium-emitting gas \citep{fransson1989late, li1993ii, Jerkstrand2012}, their coexistence suggests a range of densities within the ejecta, with both low-density regions where forbidden emission can survive and higher-density regions where the permitted \CaII{} NIR triplet remains strong. 

\begin{figure}
\includegraphics[width=1.\linewidth]{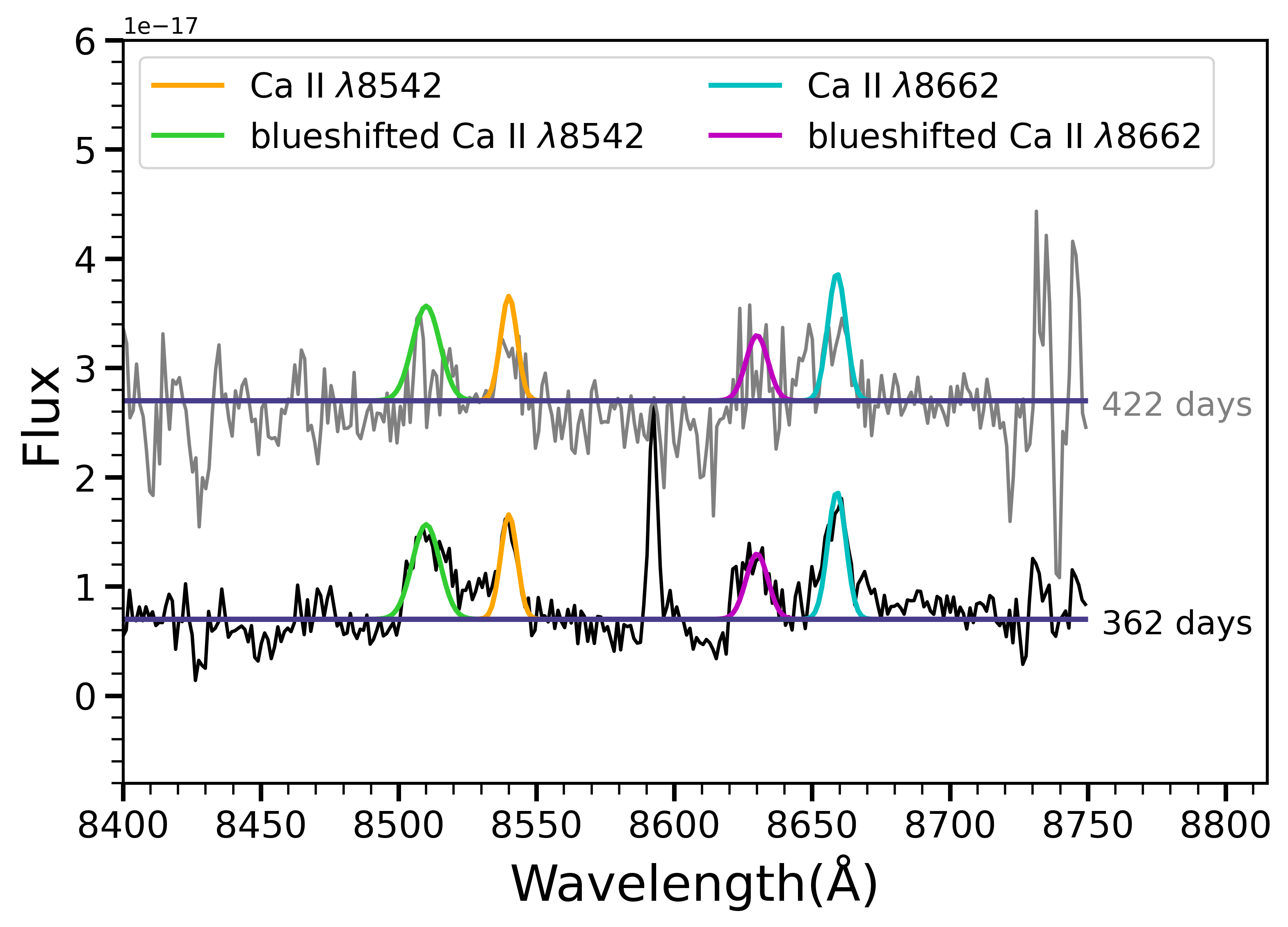}
\caption{Gaussian fits for each component of the \CaII{} near-infrared triplet. A blueshifted component for each of the detected \CaII{} lines with consistent velocity indicate the supernova origin of these lines.
\label{fig:cagaussian}}
\end{figure}

Although the \CaII{} NIR triplet features are weak at the epochs where nebular spectra were collected and therefore are not shown in Figure \ref{fig:nebs}, Gaussian decomposition of the individual components allows us to identify and characterize them. At 362 days after explosion, the \CaII{} near-infrared triplet and \Ha{} are clearly detected. The nebular emission lines are broadly centered near their rest wavelengths, but show asymmetric profiles with a prominent blueshifted component, indicating that parts of the ejecta are moving toward the observer at higher velocity and suggesting an asymmetric explosion geometry \citep{anderson2012blue}. We fit four Gaussian models to each component of the \CaII{} triplet visible ($\lambda 8542, \lambda 8662$) and their blueshifted components (Figure \ref{fig:cagaussian}). Besides the inherent change in noise between the nebular spectra from day 362 and day 422, the shape of the \CaII{} lines is similar. The blueshift of both of these components equates to a velocity of $\sim1040$ km s$^{-1}$, consistent with our other notably low measured velocities in the nebular phase. 

Residual narrow lines from the star-forming region where the SN exploded are also detected (ex. [\SII{}], [\NII{}], \Ha{}). In addition, we detect \HeI{} lines at $\lambda$4922, $\lambda$5876, $\lambda$6678, and $\lambda$7065, and very weak emission from \MgI{}] $\lambda4571$ and [\FeII{}] $\lambda7155$. 

The detected lines exhibit unusually low velocities in the nebular phase, even lower than in low-luminosity SNe IIP like SN\,2005cs (see Figure \ref{fig:velocity}). These velocities are broadly consistent with those predicted by the 9 \msun{} neutrino-driven explosion models of \cite{jerkstrand2018}, which represent the lowest-mass core-collapse progenitors explored in that framework. However, even this lowest-mass model predicts detectable [\OI{}] emission in the nebular phase, which is not observed in SN\,2023rve. Additionally, our constraints from hydrodynamic modeling, discussed in Section \ref{sec: prog_properties}, point to a more massive progenitor.

\section{Progenitor Properties}\label{sec: prog_properties}

\subsection{Nickel Mass}\label{sec:nickel}
Since radioactive decay ($\mathrm{^{56}Ni \rightarrow {^{56}Co} \rightarrow {^{56}Fe}}$) drives the nebular phase light curve decline, the late-time bolometric luminosity can be used to constrain the mass of nickel. To compute the nickel mass, we use the pseudo-bolometric light curve of  SN\,2023rve and compare it with that of SN\,1987A.


\begin{figure}
\includegraphics[width=1.\linewidth]{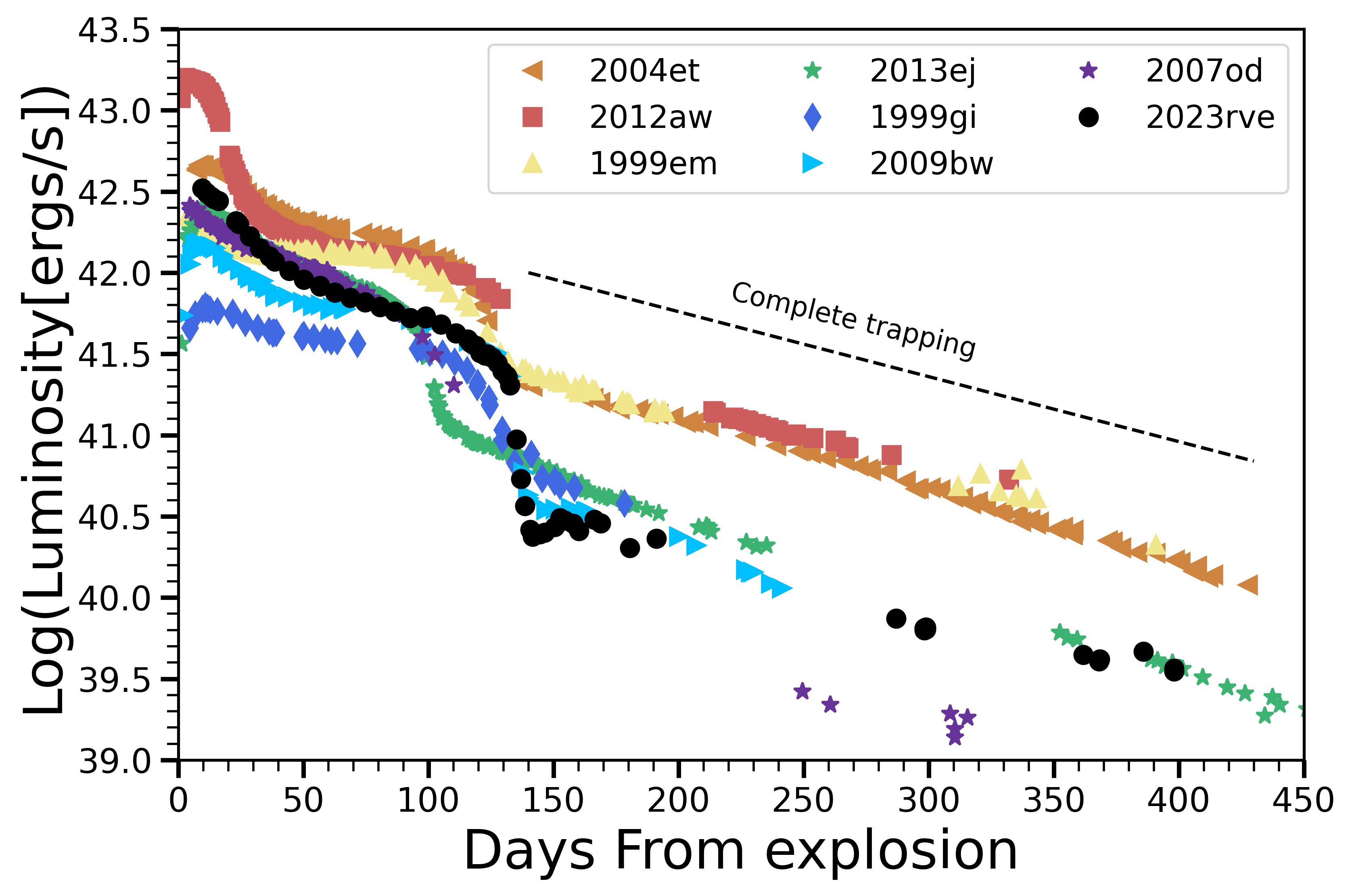}
\caption{The pseudo-bolometric light curve for SN\,2023rve and a sample of similar SNe II. The dashed line shows the decline with complete trapping, with a slope of -0.4 dex per 100 days. SN\,2023rve has tail evolution consistent with nearly complete trapping. 
\label{fig:bolo}}
\end{figure}

The pseudo-bolometric light curve of SN\,2023rve was constructed by converting the observed \textit{BVgri} magnitudes to fluxes and integrating using a quadratic polynomial, following \cite{Valenti2008}. Data from days 140 to 400 during the radioactive decay tail phase were used to calculate the pseudo-bolometric luminosity of SN\,2023rve in the {\textit{BVgri}} bands, which were compared against the pseudo-bolometric light curve of SN\,1987A using synthetic \textit{BVgri} photometry from the SNDavis database. The light curve plateau and drop-off epoch is most similar to that of SN\,1999em, SN\,2012aw, and SN\,2009bw (Figure \ref{fig:bolo}).

Assuming similar SED coverage between SN\,2023rve and SN\,1987A, the nickel mass from the pseudo-bolometric light curve comparison is given by \cite{spiro2014low}: 
\begin{equation}
    M_\mathrm{Ni}= 0.075 \ \times \ \frac{L_\mathrm{SN}}{L_\mathrm{87A}} M_\odot\
\end{equation}
where $L$ corresponds to the pseudo-bolometric luminosity for each supernova. We use a Markov Chain Monte Carlo (MCMC) fitting routine with the nickel mass and gamma-ray trapping timescale $T_0$ as free parameters, where $T_0$ characterizes the transition of the ejecta from optically thick to optically thin to gamma rays, with larger values indicating more efficient trapping. We find a best-fit trapping timescale of $T_0=304.98^{+31.44}_{-37.37}$ days, somewhat near the characteristic gamma-ray trapping timescale of 530 days adopted for SN\,1987A \citep{jerkstrandthesis}, indicating that gamma-ray trapping in SN\,2023rve is nearly complete.

From this method, we estimate a $^{56}$Ni mass of $0.0064 \pm 0.0026 \ M_\odot$, low compared to typical SNe IIP. For instance, SN\,2009bw has an established $^{56}$Ni mass of 0.022 $\ M_\odot$ \citep{2009bw}, and SN\,2012aw of 0.06 $\ M_\odot$ \citep{2012aw}.

\subsection{Progenitor Mass}\label{sec:mass}
The progenitor mass can be constrained through several different techniques, and is essential to understanding the evolution and origins of a supernova. We attempt to constrain the mass here through hydrodynamic modeling of the light curve.

\subsubsection{Hydrodynamic Modeling Constraints}
One way of constraining the progenitor mass is by comparing the SN light curve with a grid of hydrodynamic models from different progenitors. To do this, we used \texttt{SNEC} (SuperNova Explosion Code; \citealt{morozova2015light}), a one-dimensional Lagrangian hydrodynamics code that follows the evolution of the SN ejecta using input progenitor structures from a stellar evolution code (\texttt{KEPLER}; \citealt{weaver1978,Woosley2007,woosley2015,sukhboldwoosley2014,Sukhbold2016ApJ...821...38S}). To these progenitor models, we add a steady-state wind to quantify the effect of CSM on mass-loss with density profile
\[\rho(R)=\frac{\dot M}{4\pi r^2\nu_{wind}}=\frac{K}{r^2}\]
where \(K\) is the density parameter.

\texttt{SNEC} computes the bolometric light curve by modeling the hydrodynamic and radiative evolution of the expanding ejecta. Under the assumption of a blackbody photosphere, synthetic multi-band light curves can be derived and directly compared with observations. We explore a broad range of physical parameters within \texttt{SNEC}, including progenitor mass, explosion energy, circumstellar medium (CSM) density parameter, and wind extent, and compare the resulting synthetic light curves with the observations. We apply this method to characterize the light curve rise and plateau, where the assumptions of radiative diffusion and thermodynamic equilibrium are valid (out to approximately day 120).

\begin{figure}[ht]
\includegraphics[width=1.\linewidth]{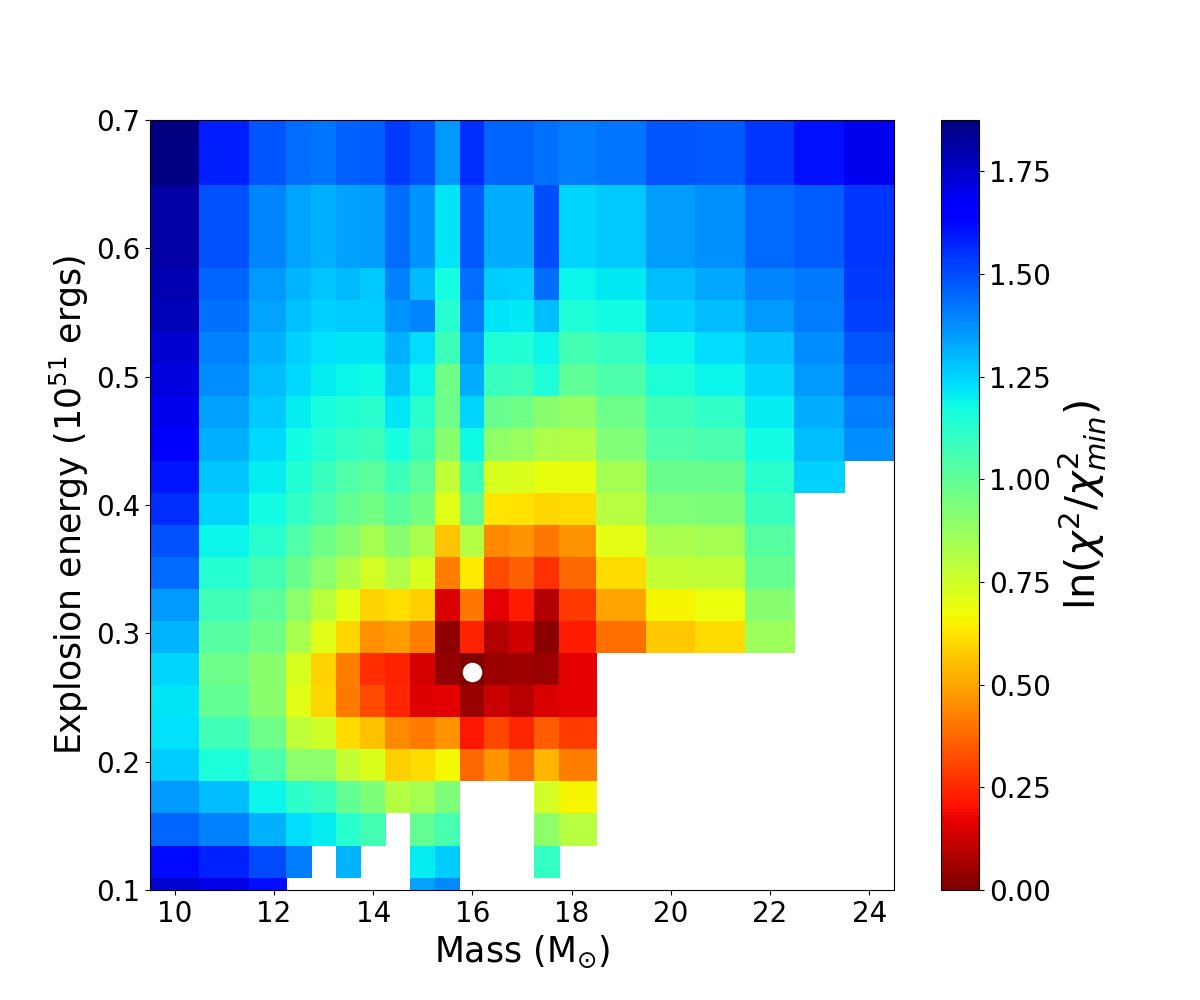}
\caption{Grid of progenitor mass and explosion energy values from \texttt{SNEC}, with color indicating the $\chi^2$ value. The lowest-$\chi^2$ parameters (16 \msun{} and 0.27$\times10^{51}$ ergs) are marked with a white dot. The missing region at high progenitor mass and low explosion energy corresponds to models that did not produce successful explosions in \texttt{SNEC}.
\label{fig:snecME}}
\end{figure}

Following the procedure laid out in \cite{morozova2017unifying}, we first varied the progenitor mass and explosion energy, running simulations for ZAMS masses between 10--24 \msun{} and explosion energies between $(0.100$--$1.000)\times10^{51}$ erg while keeping the CSM radius and density parameter fixed at zero. The nickel mass was fixed to 0.0064 \msun{} (Section \ref{sec:nickel}). We then performed a $\chi^2$ comparison between the synthetic and observed light curves over the plateau phase (40--120 days after explosion), where the effects of CSM interaction are minimal and the assumptions of \texttt{SNEC} remain valid. The nickel was mixed out to 5 \msun{} following \cite{morozova2017unifying}; this parameter does not significantly affect the plateau light curve and is not easily constrained without modeling the fall from plateau. A grid of this first-step analysis is shown in Figure \ref{fig:snecME}. The missing corner of the grid at high progenitor mass and low explosion energy corresponds to models that did not produce successful explosions in \texttt{SNEC}, likely because the explosion energy was insufficient to fully unbind the stellar envelope. 

\begin{figure}[ht]
\includegraphics[width=1.\linewidth]{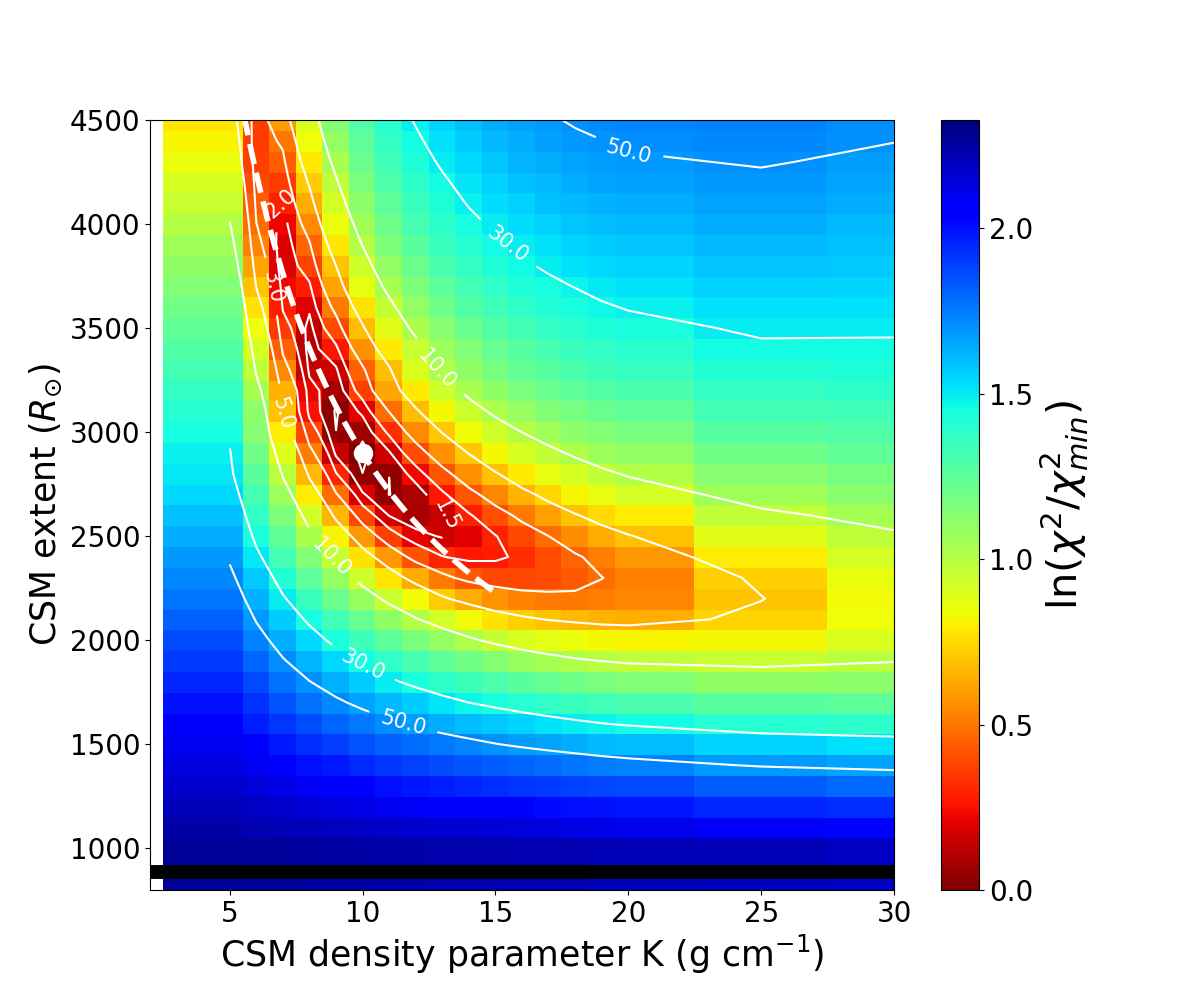}
\caption{Grid of the CSM density parameter (K) and radial extent values from \texttt{SNEC}, with color indicating the $\chi^2$ value. The lowest-$\chi^2$ parameters (1$\times10^{18}$g cm$^{-1}$ and 2900 \(R_\odot\)) are marked with a white dot. The grey line at the bottom indicates the progenitor radius without CSM (886 \(R_\odot\)), and the solid white contours show levels of constant $\chi^2/\chi^2_{\rm min}$. The dashed white line shows the line of constant CSM mass corresponding to the lowest-$\chi^2$ model, illustrating the degeneracy between K and radial extent \citep{morozova2017unifying}. The CSM mass of this best-matched model is 0.886 \msun{}. 
\label{fig:snecRK}}
\end{figure}

In the second step, we fixed the progenitor mass and explosion energy to the lowest-$\chi^2$ model from Figure \ref{fig:snecME} and varied the CSM wind extent ($R$) and density parameter ($K$). Since CSM interaction primarily affects the earliest phases of the light curve \citep{morozova2018measuring}, we restricted this comparison to the first 40 days after explosion. These results are shown in Figure \ref{fig:snecRK}. This two-step process significantly reduces the number of models needed, and allows for a more precise analysis at chosen parameters.

Although \texttt{SNEC} provides an estimate of the photospheric velocity, we did not include it in the $\chi^2$ analysis. The photospheric velocity at day 50 is tightly coupled to the luminosity ($v_\mathrm{ph,50} \propto L_{50}^{1/2}$), due to homologous expansion and a roughly fixed photospheric temperature, meaning it cannot provide an independent constraint on explosion properties \citep{Goldberg_2019}. Furthermore, while explosion energy and ejecta velocity are degenerate observables, in \texttt{SNEC} this degeneracy cannot be independently explored because the velocity structure is self-consistently determined by the input progenitor model and explosion energy. Decoupling these quantities would require modifying the underlying progenitor structures from the stellar evolution models, which is beyond the scope of this work. As a result, including the photospheric velocity does not provide additional independent constraints and does not significantly alter the lowest-$\chi^2$ parameter values.

\begin{figure}[ht]
\includegraphics[width=1.\linewidth]{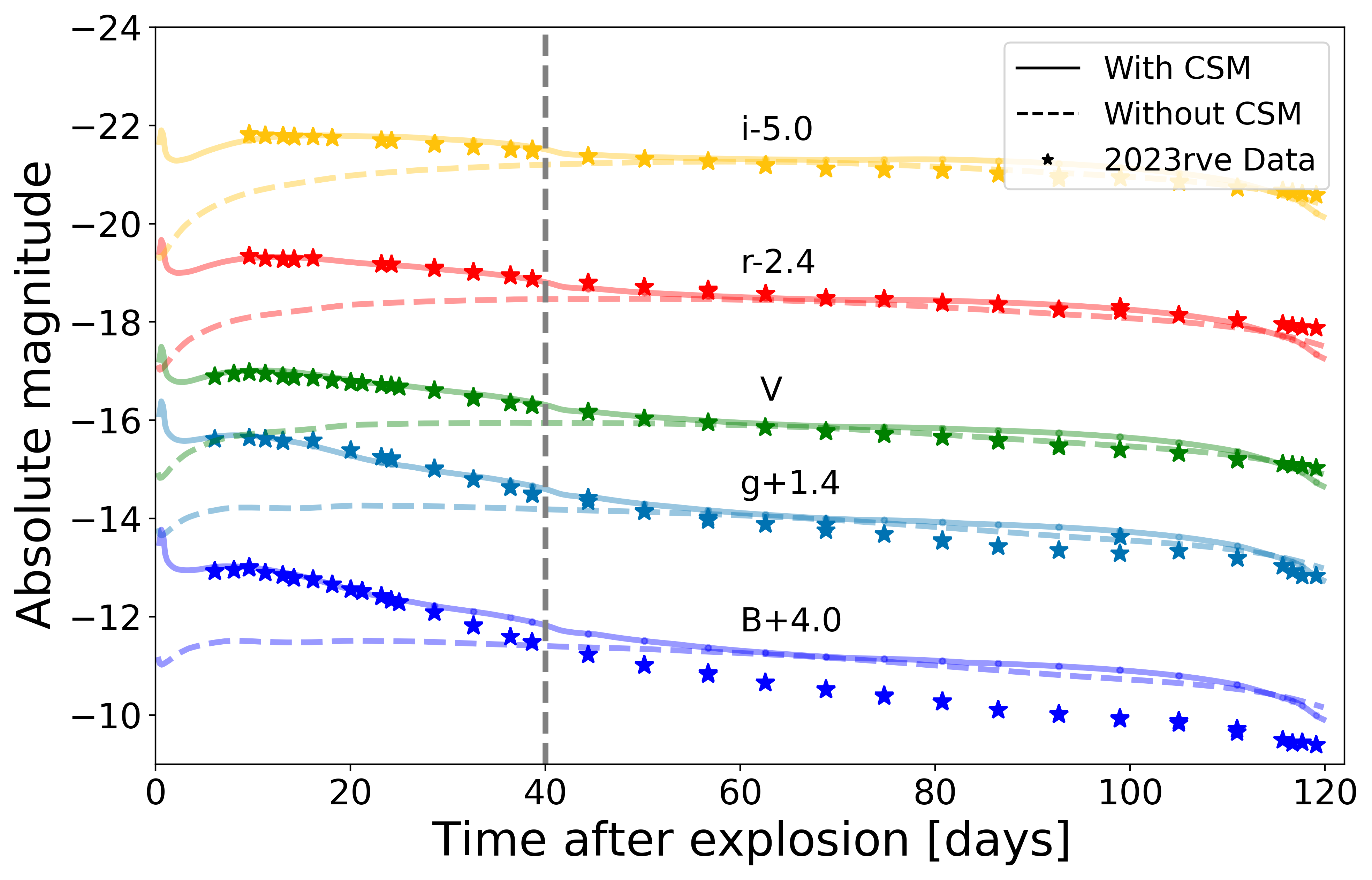}
\caption{Stars represent observational data from SN\,2023rve, the solid line shows the lowest-$\chi^2$ parameters with CSM interaction, and the dashed line shows the best-matched light curves without CSM. It is clear that the CSM is necessary to explain the behavior of the SN\,2023rve light curve at early time.
\label{fig:snecfinal}}
\end{figure}

Within the explored SNEC model grid, the lowest-$\chi^2$ model corresponds to a ZAMS mass of 16~$M_\odot$, an explosion energy of 0.27$\times10^{51}$ erg, a CSM extent of 2900~$R_\odot$, and a CSM density parameter of $1\times10^{18}$ g cm$^{-1}$. These parameters should be interpreted as the preferred solution within the adopted model framework rather than as unique constraints on the progenitor. Additional degeneracies related to progenitor structure, stellar evolution prescriptions, progenitor radius, and physical processes not included in the calculations (e.g., fallback accretion) are not explored in the present work.
This mass corresponds to a progenitor radius of approximately 886 $R_\odot$, according to stellar evolution models of \cite{Sukhbold2016ApJ...821...38S}. This radius is not independently constrained by the light-curve fitting, but instead follows directly from the adopted progenitor model, where each ZAMS mass corresponds to a unique stellar radius. A comparison of the model with these values and the observed photometry is shown in Figure \ref{fig:snecfinal}.

The discrepancy between the model and observations is largest in the $B$ band, which is expected given that \texttt{SNEC} assumes a blackbody photosphere. At late times and at blue wavelengths, this approximation becomes less valid as line blanketing removes flux from the blue, causing the model to overestimate the observed $B$-band luminosity \citep{kasen2009}.

\section{Discussion and Conclusions}\label{sec:conclusions}
In this paper, we have shown spectroscopic and photometric data for SN\,2023rve in the galaxy NGC 1097. 
This object reached a peak V-band absolute magnitude of -17.1 mag, making it relatively bright in comparison with other SNe II. It showed a plateau duration of $\sim135$ days with an early decline rate of 0.9 mag per 50 days, and showed low expansion velocities throughout its evolution. The spectroscopic evolution shows a transition from a blue continuum with broad, shallow \Ha{} at early times to metal-dominated features including \FeII{}, \ScII{}, and \BaII{} by day 30, consistent with a cooling photosphere.  Overall, SN\,2023rve exhibits a long plateau together with a rapid decline into the radioactive tail phase. While the early decline resembles the behavior of some Type IIL SNe, we interpret it as primarily due to interaction of the SN ejecta with CSM. Features like Cachito (high-velocity H$\alpha$) provide evidence of this CSM interaction with the SN ejecta, which is corroborated by the comparison of the light curve with the hydrodynamic model computed with \texttt{SNEC}. We caution that the dense circumstellar material inferred from our models should not necessarily be interpreted as evidence for enhanced pre-supernova mass loss, since red supergiants can possess extended atmospheric structures that may produce similar observational signatures \citep{2025Galax..13...72V}.

The low mass of nickel synthesized (0.0064 \msun{}) was found using the pseudo-bolometric light curves of SN\,2023rve and SN\,1987A. 
It was used with SNEC for hydrodynamic modeling of the light curve. Within the explored grid of models, the preferred solution corresponds to a progenitor mass $\sim$ 16~$M_\odot$, an explosion energy of 0.27$\times10^{51}$ erg, a CSM radius of 2900~$R_\odot$, and a CSM density parameter of $1\times10^{18}$ g cm$^{-1}$. We caution that these values are model-dependent and do not represent unique constraints on the explosion properties.
The EPM (expanding photosphere method) produced a distance consistent with published Tully-Fisher measurements and independently verified the explosion epoch.

Contrary to what we would expect from a massive progenitor star, we do not observe any [\OI{}] emission in the nebular spectra of SN\,2023rve, which were taken out to 485 days after explosion. Based on our archival comparison sample of 26 Type II SNe, the absence of nebular [\OI{}] emission seen in SN\,2023rve appears to be rare, with only two other known objects missing [\OI{}] on the same timescale. The absence of nebular [\OI{}] emission from a $\sim$14-18 \msun{} progenitor therefore remains puzzling. 

As proposed in Section \ref{sec:nebular}, a possible explanation is that SN\,2023rve is a fallback supernova, where part of the ejecta fails to reach escape velocity and falls back onto the newly-formed compact remnant. This concept was first proposed by \cite{colgate1971} and has since been invoked to explain the unusual behavior of many peculiar supernovae. 

In particular, fallback SNe should produce a low-energy explosion, and produce very little nickel and weak or absent \FeII{}, [\OI{}], and \MgI{}] lines \citep{2009ApJ...707..193F, zampieri2003}. SN\,2023rve shares several behaviors with fallback SNe and faint Type IIP SNe: nebular spectra showing absent [\OI{}] and weak \MgI{}] and [\FeII{}], low expansion velocities, and a steep decline from a long plateau into a faint radioactive tail. Low luminosity SNe IIP have themselves been proposed to be related to fallback explosions \cite{zampieri2003}. However, the direct detections of some low-luminosity Type IIP SNe progenitors \citep{smartt2009progenitors} and more recent studies (e.g.,\cite{das2026lliip,Lisakov:2017uue}) have suggested that low-luminosity SNe IIP are more likely related to low-mass 8-10 \Msun{} progenitors. Furthermore, the fallback scenario is more difficult to reconcile with the other two SNe II lacking nebular oxygen, SN\,2007od and SN\,2015C, which are both higher-velocity objects and would require different physical conditions to produce fallback. 

The absence of nebular [\OI{}] emission may also be due to other physical processes. Cooling through CO molecule formation can lower the temperature of the inner ejecta and suppress typical oxygen line emission \citep{Dessart_2025}. Dust formation at late times could occur in the cooled regions, further reducing oxygen emission by absorbing or scattering photons \citep{bostroem2025}; interestingly, both SN\,2023rve and SN\,2007od show evidence of dust formation in their ejecta \citep{2007odandrews}, suggesting this may be a common factor among SNe II lacking nebular oxygen emission. Additionally, if oxygen is well mixed with calcium in the inner ejecta, calcium becomes the dominant coolant and reduces the [\OI{}] flux \citep{Dessart2021}, which may be consistent with the [\CaII{}] emission detected in our nebular spectra. Interaction with a dense CSM shell may affect the optical thickness and ionization state of the inner ejecta, which could also potentially suppress the formation of typical oxygen emission lines. 

We also detect helium lines with intermediate FWHM values: broader than host-galaxy features but slightly narrower than other SN lines. However, their intensity does not decline with time in the same way as the supernova features, leaving their origin ambiguous. 

SN\,2023rve highlights the importance of obtaining nebular-phase spectroscopy for large samples of Type II supernovae. Future surveys in the Rubin Observatory LSST era will likely uncover additional rare events with unusual nebular properties, making systematic late-time spectroscopic follow-up essential for understanding the diversity of core-collapse supernova explosions.

\section*{Acknowledgments}

We would like to thank the anonymous reviewer for constructive suggestions that helped to improve this manuscript. We additionally thank G\'eza Cs\"ornyei for useful comments and feedback.
Time-domain research by the UC Davis team and S.V. is supported by U.S. National Science Foundation (NSF) grant AST-2407565.
Time domain research by the University of Arizona team and
D.J.S. is supported by NSF grants 2308181,
2407566, and 2432036. Supernova research at Rutgers University is supported in part by NSF award AST-2407567.
This work makes use of observations from the Las Cumbres Observatory network.
The LCO team is supported by NSF grants AST-2308113 and AST-1911151.
KAB is supported by an LSST-DA Catalyst Fellowship; this publication was thus made possible through the support of Grant 62192 from the John Templeton Foundation to LSST-DA.

Some of the data presented in this paper were obtained at the W. M. Keck  Observatory, which is operated as a scientific partnership among the California Institute of Technology, the University of California, and NASA; the observatory was made possible by the generous financial support of the W. M. Keck Foundation.

This research made use of the NASA/IPAC Extragalactic Database (NED, https://ned.ipac.caltech.edu/), which is funded by the National Aeronautics and Space Administration and operated by the California Institute of Technology.

%

\facilities{ADS, DLT40 (Prompt5, Prompt-MO), Lick (KAIT, Nickel), ATLAS, LCOGT (SBIG, Sinistro, FLOYDS), Keck:I (LRIS), Keck:II (DEIMOS, NIRES), LCOGT (Sinistro), NED, SALT (RSS), Shane (Kast)}


\software{Astropy \citep{astropy13,astropy18}, 
          HOTPANTS \citep{Becker2015},
          Matplotlib \citep{Hunter2007},
          NumPy \citep{2020Natur.585..357H},
          PYRAF \citep{2012ascl.soft07011S},
          Pandas \citep{mckinney-proc-scipy-2010},
          SciPy \citep{2020NatMe..17..261V},
          SWarp \citep{Bertin2002},
          HOTPANTS \citep{Becker2015},
          LCOGTSNpipe \citep{Valenti2016}, 
          Light Curve Fitting \citep{hosseinzadeh_light_2020},
          PypeIt \citep{pypeit:zenodo}
          }


\newpage

\appendix
\renewcommand{\thetable}{A\arabic{table}}
\setcounter{table}{0}
\vspace{-\baselineskip}
\setlength{\tabcolsep}{14pt}
\renewcommand{\arraystretch}{1.2}
\begin{deluxetable*}{lcccccc}[h]
    \tabletypesize\scriptsize
    \tablecaption{Sample of photometric observations of SN~2023rve.}
    \tablewidth{\textwidth}
    \tablehead{
    \colhead{UTC Date} &
    \colhead{MJD} &
    \colhead{Phase (days)} &
    \colhead{Magnitude} &
    \colhead{Magnitude Error} & 
    \colhead{Filter} &
    \colhead{Source}
    }
    \startdata 
    2023-09-06 & 60193.04 & 1.52 & 14.40 & 0.04 & R & Telescope Live \\
    2023-09-06 & 60193.07 & 1.55 & 14.50 & 0.04 & V & Telescope Live \\
    2023-09-10 & 60197.58 & 6.06 & 13.98 & 0.01 & U & COJ 1m \\
    2023-09-10 & 60197.58 & 6.07 & 13.98 & 0.01 & U & COJ 1m \\
    2023-09-10 & 60197.59 & 6.07 & 14.64 & 0.01 & B & COJ 1m \\
    2023-09-10 & 60197.59 & 6.08 & 14.63 & 0.01 & B & COJ 1m \\
    2023-09-10 & 60197.59 & 6.08 & 14.53 & 0.01 & V & COJ 1m \\
    2023-09-10 & 60197.59 & 6.08 & 14.52 & 0.01 & V & COJ 1m \\
    2023-09-10 & 60197.60 & 6.08 & 14.50 & 0.01 & g & COJ 1m
    \enddata
    \tablenotetext{}{(The complete table is available in machine-readable form.)}
\end{deluxetable*}\label{tab:log_of_phot}

\setlength{\tabcolsep}{20pt}
\renewcommand{\arraystretch}{1.1}


\begin{deluxetable*}{lccccc}
    \tabletypesize\scriptsize
    \tablewidth{\textwidth}
    \tablecaption{Log of spectroscopic observations of SN~2023rve.\label{tab:log_of_spec}}
    \tablehead{
    \colhead{UTC Date} &
    \colhead{MJD} &
    \colhead{Phase (days)} &
    \colhead{Telescope} &
    \colhead{Instrument} & 
    \colhead{Range (\AA)} 
    }
    \startdata 
    2023-09-10 & 60197.57 & 6.05 & FTS & FLOYDS & $3498-9999$ \\
    2023-09-10 & 60197.59 & 6.08 & BL41 & FOSC-E5535 & $4203-8092$ \\
    2023-09-11 & 60198.57 & 7.06 & FTS & FLOYDS & $3498-9998$ \\
    2023-09-12 & 60199.58 & 8.06 & FTS & FLOYDS & $3498-9997$ \\
    2023-09-13 & 60200.63 & 9.11 & FTS & FLOYDS & $3497-9998$ \\
    2023-09-14 & 60201.65 & 10.13 & FTS & FLOYDS & $3497-9998$ \\
    2023-09-15 & 60202.65 & 11.14 & FTS & FLOYDS & $3498-9997$ \\
    2023-09-16 & 60203.66 & 12.15 & FTS & FLOYDS & $3497-9998$ \\
    2023-09-17 & 60204.67 & 13.16 & FTS & FLOYDS & $3497-9997$ \\
    2023-09-18 & 60205.68 & 14.16 & FTS & FLOYDS & $3498-9998$ \\
    2023-09-19 & 60206.29 & 14.77 & VLT & FORS2 & $3379-9634$ \\
    2023-09-19 & 60206.68 & 15.17 & FTS & FLOYDS & $3497-9999$ \\
    2023-09-20 & 60207.75 & 16.24 & FTS & FLOYDS & $3497-9997$ \\
    2023-09-21 & 60208.37 & 16.86 & VLT & FORS2 & $3379-9634$ \\
    2023-09-22 & 60209.56 & 18.04 & FTS & FLOYDS & $3497-9999$ \\
    2023-09-23 & 60210.59 & 19.07 & FTS & FLOYDS & $3497-9998$ \\
    2023-09-24 & 60211.61 & 20.1 & FTS & FLOYDS & $3497-9997$ \\
    2023-09-25 & 60212.63 & 21.12 & FTS & FLOYDS & $3497-9999$ \\
    2023-09-27 & 60214.64 & 23.13 & FTS & FLOYDS & $3497-9997$ \\
    2023-09-28 & 60215.76 & 24.24 & FTS & FLOYDS & $3498-9999$ \\
    2023-09-29 & 60216.59 & 25.08 & FTS & FLOYDS & $3498-9999$ \\
    2023-10-03 & 60220.50 & 28.98 & FTN & FLOYDS & $3300-9998$ \\
    2023-10-07 & 60224.67 & 33.16 & FTS & FLOYDS & $3293-9999$ \\
    2023-10-11 & 60228.48 & 36.96 & FTN & FLOYDS & $3498-9999$ \\
    2023-10-20 & 60237.46 & 45.94 & FTS & FLOYDS & $3499-9999$ \\
    2023-10-31 & 60248.42 & 56.91 & FTN & FLOYDS & $3498-9999$ \\
    2023-11-11 & 60259.67 & 68.16 & FTS & FLOYDS & $3497-9998$ \\
    2023-11-26 & 60274.35 & 82.84 & FTN & FLOYDS & $3498-9999$ \\
    2023-12-06 & 60284.58 & 93.06 & FTS & FLOYDS & $3499-9998$ \\
    2023-12-17 & 60295.60 & 104.09 & FTS & FLOYDS & $3499-9998$ \\
    2023-12-28 & 60306.45 & 114.94 & FTS & FLOYDS & $3498-9999$ \\
    2024-09-01 & 60554 & 362.48 & Keck & LRIS & $3153-10210$ \\
    2024-10-31 & 60614 & 422.48 & Keck & LRIS & $3143-10214$ \\
    2025-01-02 & 60677 & 485.48 & Keck & LRIS & $3147-7684$ \\
    2025-01-09 & 60684 & 492.48 & SALT & RSS & $3200-9000$ \\
    \enddata
    
\end{deluxetable*}



\clearpage

\vspace{-\baselineskip}
\bibliography{sn2023rve}{}
\bibliographystyle{aasjournal}



\end{CJK*} 
\end{document}